\documentclass[journal]{IEEEtran}

\usepackage{amsmath}
\usepackage{amssymb}
\usepackage{latexsym}
\usepackage{array,arydshln}
\usepackage{multirow}
\usepackage{graphicx}
\usepackage{float}
\usepackage{bm}
\usepackage{ctable}
\usepackage{cite} 
\usepackage{balance}

\newtheorem{definition}{Definition}
\newtheorem{theorem}{Theorem}
\newtheorem{lemma}{Lemma}
\newtheorem{proposition}{Proposition}
\newtheorem{example}{Example}

\newcommand{\bs}[1]{\ensuremath{\boldsymbol{#1}}}

\newcommand{\zero}{\boldsymbol{0}}

\renewcommand{\u}{\bs{u}}

\newcommand{\epsSCone}{\varepsilon^1_{\mathrm{SC}}}
\newcommand{\epsSCthree}{\varepsilon^3_{\mathrm{SC}}}
\newcommand{\epsSCfive}{\varepsilon^5_{\mathrm{SC}}}

\newcommand{\Ma}{\mathcal{M}_{\alpha}}
\newcommand{\Mb}{\mathcal{M}_{\beta}}

\newcommand{\ma}{\boldsymbol{m}_{\alpha}}
\newcommand{\mb}{\boldsymbol{m}_{\beta}}

\newcommand{\CPCCa}{\mathcal{C}_{\mathrm{PCC}}}
 
\newcommand{\CPCCSa}{\mathcal{C}_{\mathrm{SC-PCC}}}
 
\newcommand{\CBCC}{\mathcal{C}_{\mathrm{BCC}}}

\newcommand{\CSCCa}{\mathcal{C}_{\mathrm{SCC}}}
 
\newcommand{\CSCCSa}{\mathcal{C}_{\mathrm{SC-SCC}}}

\begin{document}
%
\title{Spatially Coupled Turbo-Like Codes}



\author{Saeedeh Moloudi,~\IEEEmembership{Student Member,~IEEE}, Michael Lentmaier,~\IEEEmembership{Senior Member,~IEEE},\\ and Alexandre Graell i Amat,~\IEEEmembership{Senior Member,~IEEE}
\thanks{Parts of this paper have been presented at the IEEE
  International Symposium on Information Theory 2014, the International
  Conference on Signal Processing and Communications 2014,  the
  8th International Symposium on Turbo Codes \& Iterative Information
  Processing 2014, the Eleventh International Symposium on Wireless
  Communication Systems 2014, and at the IEEE Information Theory
  Workshop 2015.}\thanks{This work was supported in part by the Swedish Research Council (VR) under grant \#621-2013-5477.}
\thanks{S. Moloudi and M. Lentmaier are with the Department of Electrical and Information Technology, Lund University, Lund, Sweden (e-mail:  \{saeedeh.moloudi,michael.lentmaier\}@eit.lth.se).}
\thanks{A. Graell i Amat is with the Department of Electrical Engineering, Chalmers University of Technology, SE-41296 Gothenburg, Sweden (e-mail: alexandre.graell@chalmers.se).}
\thanks{Copyright (c) 2017 IEEE. Personal use of this material is permitted. However, permission to use this material for any other purposes must be obtained from the IEEE by sending a request to pubs-permissions@ieee.org.}
}



\maketitle

\vspace*{-1cm}
\begin{abstract}
In this paper, we introduce the concept of spatially coupled turbo-like codes (SC-TCs) as the spatial coupling of a number of turbo-like code ensembles.
In particular, we consider the spatial coupling of parallel concatenated codes (PCCs), introduced by Berrou \emph{et al.}, and that of serially concatenated codes (SCCs), introduced by Benedetto \emph{et al.}.
Furthermore, we propose two extensions of braided convolutional codes (BCCs), a class of turbo-like codes which have an inherent spatially coupled structure, to higher coupling memories, and show that these yield improved belief propagation (BP) thresholds as compared to the original BCC ensemble.
We derive the exact density evolution (DE) equations for SC-TCs and analyze their asymptotic behavior on the binary erasure channel.
We also consider the construction of families of rate-compatible SC-TC ensembles.
Our numerical results show that threshold saturation of the belief propagation (BP) decoding threshold to the maximum a-posteriori threshold of the underlying uncoupled ensembles occurs for large enough coupling memory.
The improvement of the BP threshold is especially significant for SCCs and BCCs, whose uncoupled ensembles suffer from a poor BP threshold.
For a wide range of code rates, SC-TCs show close-to-capacity performance as the coupling memory increases.
We further give a proof of threshold saturation for SC-TC ensembles with identical component encoders.
In particular, we show that the DE of SC-TC ensembles with identical component encoders can be properly rewritten as a scalar recursion.
This allows us to define potential functions and prove threshold saturation using the proof technique recently introduced by Yedla \emph{et al.}.

\end{abstract}


\IEEEpeerreviewmaketitle

\section{Introduction}

Low-density parity-check (LDPC) convolutional codes \cite{JimenezLDPCCC}, also known as spatially coupled LDPC (SC-LDPC) codes \cite{Kudekar_ThresholdSaturation}, can be obtained from a sequence of individual LDPC block codes by distributing the edges of their Tanner graphs over several adjacent blocks \cite{LentmaierTransITOct2010}.
The resulting spatially coupled codes exhibit a \emph{threshold saturation} phenomenon, which has attracted a lot of interest in the past few years: The threshold of an iterative belief propagation (BP) decoder, obtained by density evolution (DE), can be improved to that of the optimal maximum-a-posteriori (MAP) decoder, for properly chosen parameters.
It follows from threshold saturation that it is possible to achieve capacity by spatial coupling of simple regular LDPC codes, which show a significant gap between BP and MAP threshold in the uncoupled case.
A first analytical proof of threshold saturation was given in \cite{Kudekar_ThresholdSaturation} for the binary erasure channel (BEC), considering a specific ensemble with uniform random coupling. An alternative proof based on potential functions was then presented in \cite{Yedla2012,Yedla2014,KudekarWaveLike}, which was extended from scalar recursions to vector recursions in \cite{Yedla2012vector}.
By means of vector recursions, the proof of threshold saturation can be extended to spatially coupled ensembles with structure, such as SC-LDPC codes based on protographs \cite{Mitchell_ProtographSCLDPC}.

The concept of spatial coupling is not limited to LDPC codes.
Also codes on graphs with stronger component codes can be considered.
In this case the structure of the component codes has to be taken into account in a DE analysis.
Instead of a simple check node update, a constraint node update within BP decoding of a generalized LDPC code involves an a-posteriori probability (APP) decoder applied to the associated component encoder.
In general, the input/output transfer functions of the APP decoder are multi-dimensional because the output bits of the component encoder have different protection.
For the BEC, however, it is possible to analytically derive explicit transfer functions \cite{AshikhminEXIT} by means of a Markov chain analysis of the decoder metric values in a trellis representation of the considered code \cite{LentDE_PG-LDPC}.
This technique was applied in \cite{LentDE_BBC, Lent_DE_SCGLDPC} to perform a DE analysis of braided block codes (BBCs) \cite{JimenezBBC} and other spatially coupled generalized LDPC codes.
Threshold saturation could be observed numerically in all the considered cases.
BBCs can be seen as a spatially coupled version of product codes, and are closely related to staircase codes \cite{Smith12}, which have been proposed for high-speed optical communications.
It was demonstrated in \cite{PfisterISIT12,PfisterGC13} that BBCs show excellent performance even with the iterative hard decision decoding that is proposed for such scenarios.
The recently presented spatially coupled split-component codes \cite{Truhachev_SplitComponent} demonstrate the connections between BBCs and staircase codes.

In this paper, we study codes on graphs whose constraint nodes represent convolutional codes \cite{Wiberg95, WibergPhD96,Loeliger}.
We denote such codes as turbo-like codes (TCs).
We consider three particular concatenated convolutional coding schemes: Parallel concatenated codes (PCCs) \cite{BerrouTC}, serially concatenated codes (SCCs) \cite{Benedetto98Serial}, and braided convolutional codes (BCCs) \cite{ZhangBCC}.
Our aim is to investigate the impact of spatial coupling on the BP threshold of these TCs.
For this purpose we introduce some special block-wise spatially coupled ensembles of PCCs (SC-PCCs) and SCCs (SC-SCCs) \cite{Moloudi_SCTurbo}.
In the case of BCCs, which are inherently spatially coupled, we consider the original block-wise ensemble from \cite{ZhangBCC,Moloudi_DEBCC} and generalize it to larger coupling memories.
Furthermore, we introduce a novel BCC ensemble in which not only the parity bits but also the information bits are coupled over several time instants \cite{Moloudi_SPCOM14}.


For these spatially coupled turbo-like codes (SC-TCs), we perform a threshold analysis for the BEC analogously to \cite{LentmaierTransITOct2010,LentDE_BBC, Lent_DE_SCGLDPC}.
We derive their exact DE equations from the transfer functions of the convolutional component decoders \cite{Kurkoski, tenBrinkEXITConv}, whose computation is similar to that for generalized LDPC codes in \cite{LentDE_PG-LDPC}.
In order to evaluate and compare the ensembles at different rates, we also derive DE equations for the punctured ensembles.
Using these equations, we compute BP thresholds for both coupled and uncoupled TCs \cite{Moloudi_ISTW14}  and compare them with the corresponding MAP thresholds \cite{Measson2009,Measson_Turbo}.
Our numerical results indicate that threshold saturation occurs if the coupling memory is chosen sufficiently large.
The improvement of the BP threshold is specially significant for SCCs and BCCs, whose uncoupled ensembles suffer from a poor BP threshold.
We then consider the construction of families of rate-compatible SC-TCs which achieve close-to-capacity performance for a wide range of code rates.

Motivated by the numerical results, we prove threshold saturation analytically.
We show that, by few assumptions in the ensembles of uncoupled TCs, in particular considering identical component encoders, it is possible to rewrite their DE recursions in a form that corresponds to the recursion of a scalar admissible system.
This representation allows us to apply the proof technique based on potential functions for scalar admissible systems proposed in \cite{Yedla2012,Yedla2014}, which simplifies the analysis.
For the general case, the analysis is significantly more complicated and requires the coupled vector recursion framework of \cite{Yedla2012vector}.
Finally, for the example of PCCs, we generalize the proof to non-symmetric ensembles with different component encoders by using the framework in \cite{Yedla2012vector}.

The remainder of the paper is organized as follows. 
In Section~\ref{sec:CompactGraphCC}, we introduce a compact graph representation for the trellis of a convolutional code that is amenable for a DE analysis.
Furthermore, we derive explicit input/output transfer functions of the BCJR decoder for transmission over the BEC.
Then, in Section~\ref{sec:CompactGraphTCs}, we describe uncoupled ensembles of PCCs, SCCs and BCCs by means of the compact graph representation. SC-TCs, their spatially coupled counterparts, are introduced in Section~\ref{sec:SCTCs}.
In Section~\ref{sec:DE}, we derive exact DE equations for  uncoupled and coupled ensembles of TCs.
In Section~\ref{sec:RandomP}, we consider random puncturing and derive the corresponding DE equations and analyze SC-TCs as a family of rate compatible codes.
Numerical results are presented and discussed in Section~\ref{Sec6}.
Threshold saturation, which is observed numerically in the results section, is proved analytically in Section \ref{Sec7}.  
Finally, the paper is concluded in Section~\ref{Sec8}.

\section{Compact Graph Representation and Transfer Functions of Convolutional Codes}
\label{sec:CompactGraphCC}
In this section, we introduce a graphical representation of a convolutional code, which can be seen as a compact form of its corresponding factor graph \cite{Loeliger}. 
This compact graph representation makes the illustration of SC-TCs simpler and is convenient for the DE analysis.  
We also generalize the method in \cite{Kurkoski, tenBrinkEXITConv} to derive explicit input/output transfer functions of the BCJR decoder of rate-$k/n$ convolutional codes on the BEC, which will be used in Section~\ref{sec:DE} to derive the exact DE for SC-TCs.


\subsection{Compact Graph Representation}
\begin{figure}[!t]
  \centering
    \includegraphics[width=0.8\linewidth]{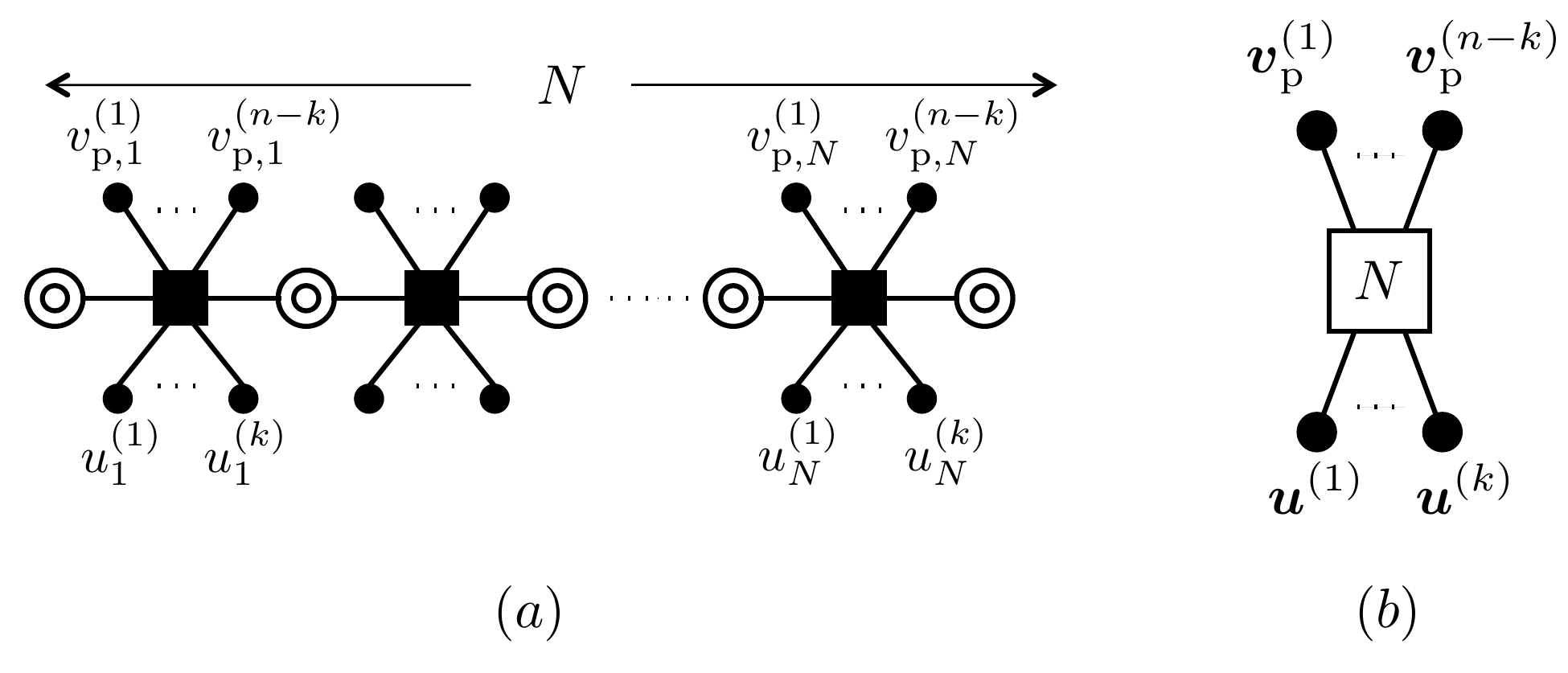}
\caption{(a) Factor graph representation of a rate-$k/n$ systematic convolutional code. (b) Compact graph representation of the same code.}
\label{factorgraph}
\end{figure}

Consider a rate-$k/n$ systematic convolutional encoder of code length $nN$ bits, i.e., its corresponding trellis has $N$ trellis sections. At each time instant $\tau=1,\ldots,N$, corresponding to a trellis section, the encoder encodes $k$ input bits and generates $n-k$ parity bits.
Let $\bs{u}^{(i)}=(u_{1}^{(i)}, u_{2}^{(i)}, \dots, u_{N}^{(i)})$, $i=1,\dots,k$, and $\bs{v}_{\text{p}}^{(i)}=(v_{\text{p},1}^{(i)}, v_{\text{p},2}^{(i)}, \dots, v_{\text{p},N}^{(i)})$, $i=1,\dots,n-k$, denote the $k$ input sequences and the $n-k$ parity sequences, respectively. We also denote by $\bs{v}^{(i)}=(v_{1}^{(i)}, v_{2}^{(i)}, \dots, v_{N}^{(i)})$, $i=1,\dots,n$, the $i$th code sequence, with $\bs{v} ^{(i)}=\bs{u}^{(i)}$ for $i=1,\dots,k$ and $\bs{v} ^{(i)}=\bs{v}_{\text{p}}^{(i-k)}$ for $i=k+1,\dots,n$. The conventional factor graph of a convolutional encoder is shown in Fig.~\ref{factorgraph}(a), where black circles represent code bits, each black square corresponds to the code constraints (allowed combinations of input state, input bits, output bits, and output state) of one trellis section, and the double circles are (hidden) state variable nodes. 

For convenience, we will represent a convolutional encoder with the more compact graph representation depicted in Fig.~\ref{factorgraph}(b). 
In this compact graph representation, each input sequence $\bs{u}^{(i)}$ and each parity sequence $\bs{v}^{(i)}_{\text{p}}$ is represented by a single black circle, referred to as variable node, i.e., each circle represents $N$ bits. Furthermore, the code trellis is represented by a single empty square, referred to as factor node. The factor node is labeled by the length $N$ of the trellis. 
Each node in the compact graph represents a sequence of nodes belonging to the same type, similar to the nodes in a protograph of an LDPC code.
Variable nodes in the original factor graph may represent different bit values, even if they belong to the same type in the compact graph.
However, assuming a tailbiting trellis, the probability distribution of these values after decoding will be equal for all variables that correspond to the same node type.
As a consequence, a DE analysis can be performed in the compact graph, independently of the trellis length $N$, which plays a similar role as the lifting factor of a protograph ensemble.
If a terminated convolutional encoder, which starts and ends in the zero state, is used instead, the bits that are close to the start and end of the trellis will have a slightly stronger protection. Since this effect will not have a significant impact on the performance, we will neglect this throughout this paper and assume equal output distributions for all bits of the trellis, even when termination is used.

\subsection{Transfer Function of the BCJR Decoder of a  Convolutional Code} \label{TransferFunction}

Consider the BCJR decoder of a memory $\nu$, rate$-k/n$ convolutional encoder and transmission over the BEC. 
Without loss of generality, we restrict ourselves within this paper to encoders with $k=1$ or $n-k=1$, which can be implemented with $2^\nu$ states in controller canonical form or observer canonical form, respectively. 
We would like to characterize the transfer function between the input erasure probabilities (i.e., prior to decoding) and output erasure probabilities (i.e., after decoding) on both the input bits and the output bits of the convolutional encoder. Note that the erasure probabilities at the input of the decoder depend on both the channel erasure probability and the a-priori erasure probabilities on systematic and parity bits (provided, for example, by another decoder). Thus, in the more general case, we consider non-equal erasure probabilities at the input of the decoder.

Consider the extrinsic erasure probability of the $l$th code bit, $l=1,2, \dots, n$, which is the erasure probability of the $l$th code bit when it is estimated based on the other code bits\footnote{Without loss of generality we assume that the first $k$ bits are the systematic bits.}. This extrinsic erasure probability,  at the output of the decoder, is denoted by $p_{l}^{\text {ext}}$. The probabilities $p_{l}^{\text {ext}}$ depend on the erasure probabilities of all code bits (systematic and parity) at the input of the decoder, 
\begin{align}
\label{eq:Transfer1}
p_ {l}^{\text {ext}}=f_l(p_1,p_2, \dots,p_n),
\end{align}
where $p_l$ is the erasure probability of the $l$th code bit at the input of the decoder and $f_l(p_1,p_2, \dots,p_n)$ is the transfer function of the BCJR decoder for the $l$th code bit. For notational simplicity, we will often omit the argument of $f_l(p_1,p_2, \dots,p_n)$ and write simply $f_l$.


Let $\bs{r}^{(i)}=(r_{1}^{(i)}, r_{2}^{(i)}, \dots, r_{N}^{(i)})$,
$i=1,\dots,n$, be the vectors of received symbols at the output of the channel, with $r_{j}^{(i)}\in\{0,1,?\}$, where $?$ denotes an erasure. The branch metric of the trellis edge departing from  state $\sigma'$ at time $\tau-1$ and ending to state $\sigma$ at time $\tau$, $\tau=1,\dots,N$,  is

\begin{align}
\gamma_{\tau}(\sigma',\sigma)=\prod_{l=1}^{n}  p\left(r_{\tau}^{(l)}\; \big| \; v_{\tau}^{(l)}\right)\cdot p\left(v_{\tau}^{(l)}\right),
\end{align}
where $p\big(v_{\tau}^{(l)}\big)$ is the a-priori probability on symbol $v_{\tau}^{(l)}$.

The forward and backward metrics of the BCJR decoder are\begin{align}
&\alpha_{\tau}(\sigma)=\sum_{\sigma'}
\gamma_{\tau}(\sigma',\sigma) \cdot \alpha_{\tau-1}(\sigma')\\
&\beta_{\tau-1}(\sigma')=\sum _{\sigma}\gamma_{\tau}(\sigma',\sigma)
\cdot \beta_{\tau}(\sigma').
\end{align}

Finally, the extrinsic output likelihood ratio is given by 
\begin{align*}
&L^{(l)}_{\text{out},\tau}=\\
&\frac{\sum\limits_{(\sigma',\sigma):v_{\tau}^{(l)}=0}\alpha_{\tau-1}(\sigma')\cdot
\gamma_{\tau}(\sigma',\sigma)\cdot
\beta_{\tau}(\sigma)}{\sum\limits_{(\sigma',\sigma):v_{\tau}^{(l)}=1}
\big(\alpha_{\tau-1}(\sigma')\cdot \gamma_{\tau}(\sigma',\sigma)\cdot \beta_{\tau}(\sigma)}\cdot\frac{p\left(v_{\tau}^{(l)}=1\right)}{p\left(v_{\tau}^{(l)}=0\right)}.
\end{align*}

Let the $2^\nu$ trellis states be $s_1,s_2,\ldots,s_{2^\nu}$. Then, we define the forward and backward metric vectors as $\boldsymbol{\alpha}_{\tau}=(\alpha_{\tau}(s_1),\ldots,\alpha_{\tau}(s_{2^\nu}))$ and $\boldsymbol{\beta}_{\tau}=(\beta_{\tau}(s_1),\ldots,\beta_{\tau}(s_{2^\nu}))$, respectively. 
For transmission on the BEC, the nonzero entries of vectors $\boldsymbol{\alpha}_{\tau}$ and $\boldsymbol{\beta}_{\tau}$ are all equal. Thus, we can normalize them to $1$.


We consider transmission of the all-zero codeword. The sets of values that vectors $\boldsymbol{\alpha}_{\tau}$ and
 $\boldsymbol{\beta}_{\tau}$ can take on are denoted by $\Ma=\{\ma^{(1)},\ldots,\ma^{(|\Ma|)}\}$ and $\Mb=\{\mb^{(1)},\ldots,\mb^{(|\Mb|)}\}$, respectively. It is important to remark that these sets are finite. Furthermore, the sequence $\dots,\boldsymbol{\alpha}_{\tau-1},\boldsymbol{\alpha}_{\tau},\boldsymbol{\alpha}_{\tau+1},\dots$
forms a Markov chain, which can be properly described by a probability transition matrix, denoted by $\boldsymbol{M}_{\alpha}$.
The $(i,j)$ entry of $\boldsymbol{M}_{\alpha}$ is the probability of transition from state
$\ma^{(i)}$ to state $\ma^{(j)}$. Denote the steady
state distribution vector of the Markov chain by $\boldsymbol{\pi}_{\alpha}$, which can be computed as the solution to
\begin{equation}
\boldsymbol{\pi}_{\alpha}=\boldsymbol{M}_{\alpha}\cdot \boldsymbol{\pi}_{\alpha}.  
\end{equation}
Similarly, we can define the transition matrix for the sequence of backward metrics $\dots,\boldsymbol{\beta}_{\tau+1},\boldsymbol{\beta}_{\tau},$ $\boldsymbol{\beta}_{\tau-1},\dots$, denoted by $\boldsymbol{M}_{\beta}$, and compute the steady state distribution vector $\bs{\pi}_{\beta}$.

\begin{example}
Consider the rate-$2/3$, $4$-state convolutional encoder with generator
matrix 
\begin{equation*}
\boldsymbol{G} (D)=\left( \begin{array}{ccc}1&0&\frac{1}{1+D+D^2}\\0&1&\frac{1+D^2}{1+D+D^2}\end{array}\right).
\end{equation*}
$\mathcal{M}_{\alpha}$
and $\mathcal{M}_{\beta}$ are equal and have cardinality $5$,
\begin{align*}
&\mathcal{M}_{\alpha}=\mathcal{M}_{\beta}=\\
&\{(1,0,0,0),(1,1,0,0),(1,0,0,1),(1,0,1,0),(1,1,1,1)\}.
\end{align*}

Consider equal erasure probability for all code bits at the input of the decoder, i.e., $p_1=p_2=p_3 = p$. Then,
\begin{align*}
\resizebox{\hsize}{!}{$
\boldsymbol{M}_{\alpha}=\left[\begin{array}{ccccc}
(1-p)^2(2p+1)&(1-p)^2&(1-p)^3&0&0\\
p^2(1-p)&0&p(1-p)^2&p^3-2p+1&(1-p)^2\\
p^2(1-p)&p(1-p)&p(1-p)^2&0&0\\
p^2(1-p)&p(1-p)&p(1-p)^2&0&0\\
p^3&p^2&p^2(3-2p)&p^2(2-p)&p(2-p)
\end{array}\right] \; .
$ 
}
\end{align*} \hfill $\triangle$
\end{example}
\vspace{3mm}

In order to compute the erasure probability of the $l$th bit at the output of the decoder, we have to
compute the probability of $L^{(l)}_{\text{out},\tau}=1$. Define the 
matrices $\boldsymbol{T}_{l}$, $l=1,2,\dots,n$, where the $(i,j)$ entry of $\boldsymbol{T}_{l}$ is computed as
\[
T_{l}(i,j)=p\left(L^{(l)}_{\text{out},\tau}=1\; | \;\boldsymbol{\alpha}_{\tau}=\ma^{(i)},\boldsymbol{\beta}_{\tau+1}=\mb^{(j)}\right).
\]
Thenhresho, the extrinsic erasure probability of the $l$th output, $p_ {l}^{\text{ext}}$, introduced in~\eqref{eq:Transfer1},  is obtained as
\begin{align}
p_ {l}^{\text {ext}}&=f_l(p_1,p_2,\dots,p_n)=p\left(L^{(l)}_{\text{out},\tau}=1\right) \nonumber\\
&=\sum^{|\mathcal{M}_{\alpha}|}_{i=1}\sum^{|\mathcal{M}_{\beta}|}_{j=1}p\left(
L^{(l)}_{\text{out},\tau}=1\; | \; \boldsymbol{\alpha}_{\tau}=\ma^{(i)},\boldsymbol{\beta}_{\tau+1}=\mb^{(j)}\right)\nonumber\\ 
&\quad\quad\quad\quad\quad \cdot p\left(\boldsymbol{\alpha}_{\tau}=\ma^{(i)}\right)\cdot p\left(\boldsymbol{\beta}_{\tau+1}=\mb^{(j)}\right)\nonumber\\
& =\boldsymbol{\pi}_{\alpha} \cdot\boldsymbol{T}_{l}\cdot \boldsymbol{\pi}_{\beta} . 
\end{align}

\begin{example}
Consider the rate$-2/3$ convolutional encoder with generator matrix
\begin{equation*}
\boldsymbol{G} (D)=\left( \begin{array}{ccc}1&0&\frac{1}{1+D}\\0&1&\frac{D}{1+D}\end{array}\right).
\end{equation*}

Assuming $p_1=p_2=p_3\triangleq p$, the transfer functions for the corresponding decoder are
\[
f_1=f_2=\frac{p(p^5-4p^4+6p^3-5p^2+2p+1)}{p^6-4p^5+6p^4-6p^3+5p^2-2p+1},
\]
\[
f_3=\frac{p^2(p^2-4p+4)}{p^6-4p^5+6p^4-6p^3+5p^2-2p+1}.
\] \hfill $\triangle$
\end{example}

\begin{lemma}
\label{Lemma1}
Consider a terminated convolutional encoder where all distinct input
sequences have distinct encoded sequences. 
For such a system, the transfer function $f(p_1,p_2,\dots,p_n)$ of a BCJR decoder with input erasure probabilities $p_1,p_2,\dots,p_n$, or any convex combination of such transfer functions, is increasing in all its arguments.
\end{lemma}
\begin{IEEEproof}
We prove the statement by contradiction. Recall that the BCJR decoder
is an optimal APP decoder. 
Now, consider the transmission of the same
codeword over two channels, called channel 1 and 2. The erasure probabilities of the $i$th bit
at the input of the decoder are denoted by $p_i^{(1)}$ and $p_i^{(2)}$ for
transmission over channel 1 and 2, respectively. These erasure probabilities are equal for all
$i=1,\ldots,n$ except for the $j$th bit, for which
$p_j^{(1)}<p_j^{(2)}$.
Assume that the transfer function $f$ is non-increasing
in its $j$th argument,
 \begin{equation}
\label{contradiction}
f(p_1^{(1)},\dots,p_j^{(1)},\ldots, p_n^{(1)}) \geq f(p_1^{(2)},\ldots,p_j^{(2)},\ldots,p_n^{(2)}).
 \end{equation}
Puncture the $j$th bit sequence of the codeword transmitted over channel 1 such that $p_{j,\text{punc}}^{(1)}=p_j^{(2)}$. Since puncturing can only make the output of the decoder worse (otherwise we could replace our encoder with the punctured one and achieve a higher  rate),
\begin{align}
\label{eq:Contr2}
f(p_1^{(1)},\ldots,p_{j,\text{punc}}^{(1)},\ldots,p_n^{(1)})&>f(p_1^{(1)},\dots,p_j^{(1)},\ldots,p_n^{(1)}),
\end{align}
Since after puncturing $p_i^{(1)}$ and $p_i^{(2)}$ are equal for all $i$, then 
$f(p_1^{(1)},\dots,p_{j,\text{punc}}^{(1)},\ldots,p_n^{(1)}) =f(p_1^{(2)},\dots,p_j^{(2)},\ldots,p_n^{(2)})$. Then, we can rewrite the inequality \eqref{eq:Contr2} as
\begin{align}
\label{eq:Contr3}
f(p_1^{(2)},\ldots,p_j^{(2)},\ldots,p_n^{(2)})&>f(p_1^{(1)},\dots,p_j^{(1)},\ldots,p_n^{(1)}).
\end{align}
However, the inequality \eqref{eq:Contr3} is in contradiction with \eqref{contradiction}.
\end{IEEEproof}

\section{Compact Graph Representation of\\ Uncoupled Turbo-like Codes}
\label{sec:CompactGraphTCs}


In this section, we describe PCCs, SCCs and BCCs using the compact graph representation introduced in the previous section. In Section~\ref{sec:SCTCs} we then introduce the corresponding spatially coupled ensembles.

\subsection{Parallel Concatenated Codes}

\begin{figure}[t]
  \centering
    \includegraphics[width=0.75\linewidth]{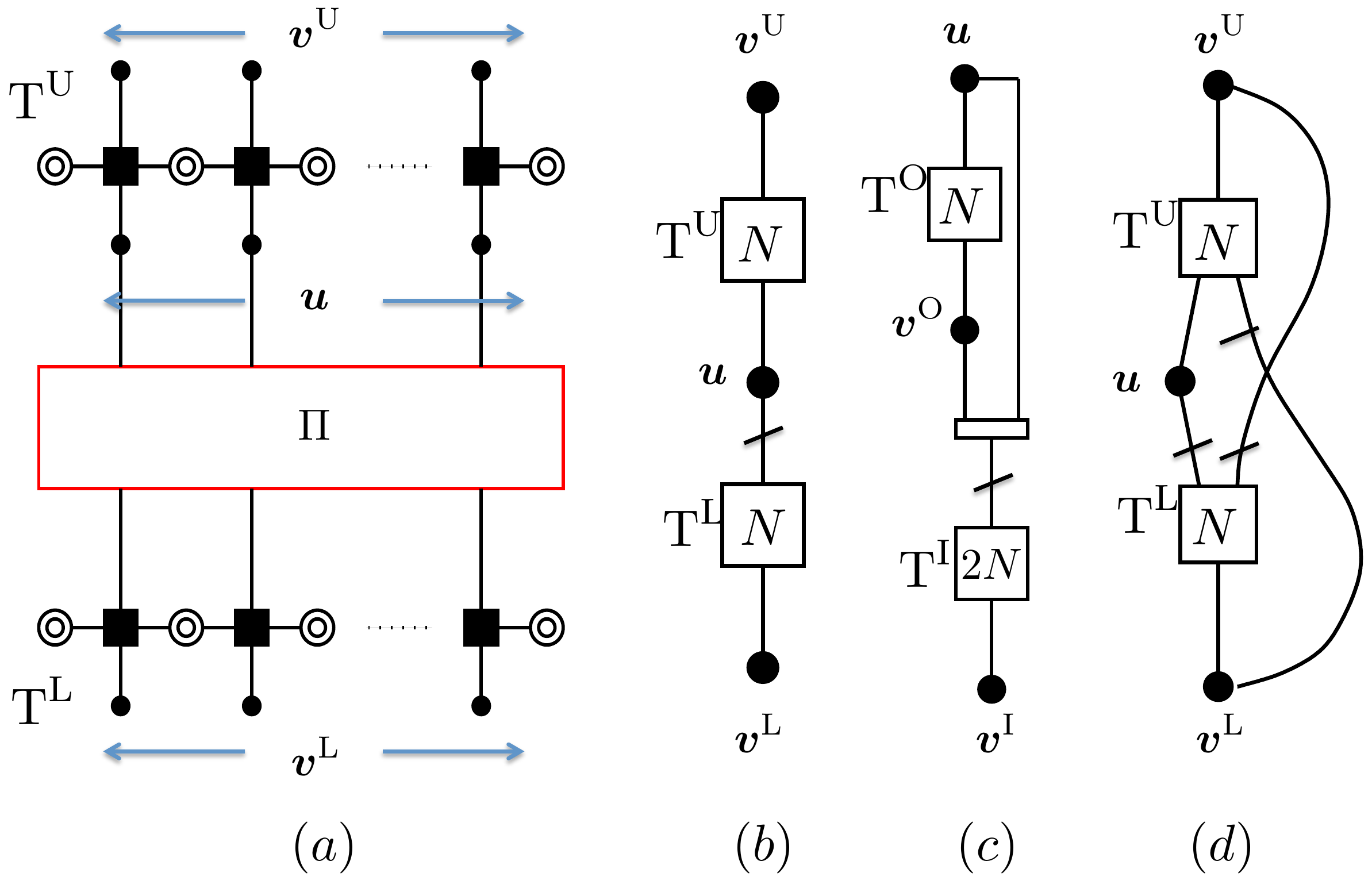}
\caption{(a) Conventional factor graph of a PCC. Compact graph representation of a (b) PCC, (c) SCC, (d) BCC.}
\label{Uncoupled}
\end{figure}

We consider a  rate $R=1/3$ PCC built from two rate-$1/2$ recursive systematic convolutional encoders, referred to as the upper and lower component encoder. 
Its conventional factor graph is shown in Fig.~\ref{Uncoupled}(a), where $\Pi$ denotes the permutation.
The trellises corresponding to the upper and lower encoders are denoted by $\text{T}^{\text{U}}$ and $\text{T}^{\text{L}}$, respectively. The information sequence $\bs{u}$, of length $N$ bits, and a reordered copy are encoded by the upper and lower encoder, respectively, to produce the parity sequences $\bs{v}^{\text{U}}$ and $\bs{v}^{\text{L}}$. The code sequence is denoted by $\bs{v}=(\bs{u},\bs{v}^{\text{U}},\bs{v}^{\text{L}})$. The compact graph representation of the PCC is shown in Fig.~\ref{Uncoupled}(b), where
each of the sequences $\bs{u}$, $\bs{v}^{\text{U}}$ and $\bs{v}^{\text{L}}$ is represented by a single variable node and the trellises are replaced by factor nodes $\text{T}^{\text{U}}$ and $\text{T}^{\text{L}}$ (cf. Fig.~\ref{factorgraph}).
In order to emphasize that a reordered copy of the input sequence is used in $\text{T}^{\text{L}}$, the permutation is depicted by a line that crosses the edge which connects $\bs{u}$ to $\text{T}^{\text{L}}$.

\subsection{Serially Concatenated Codes}

We consider a rate $R=1/4$ SCC built from the serial concatenation of two rate-$1/2$ recursive systematic component encoders, referred to as the outer and inner component encoder. Its compact graph representation is shown in Fig.~\ref{Uncoupled}(c), where $\text{T}^{\text{O}}$ and $\text{T}^{\text{I}}$ are the factor nodes corresponding to the outer and inner encoder, respectively, and the rectangle illustrates a multiplexer/demultiplexer. The information sequence $\bs{u}$, of length $N$, is encoded by the outer encoder to produce the parity sequence $\bs{v}^{\text{O}}$. Then, the sequences $\bs{u}$ and $\bs{v}^{\text{O}}$ are multiplexed and reordered to create the intermediate sequence $\tilde{\bs{v}}^{\text{O}}$, of length $2N$ (not shown in the graph). Finally, $\tilde{\bs{v}}^{\text{O}}$ is encoded by the inner encoder to produce the parity sequence $\bs{v}^{\text{I}}$. The transmitted sequence is $\bs{v}=(\bs{u},\bs{v}^{\text{O}},\bs{v}^{\text{I}})$. 

  \begin{figure*}[t]
	\centering
	\includegraphics[width=\linewidth]{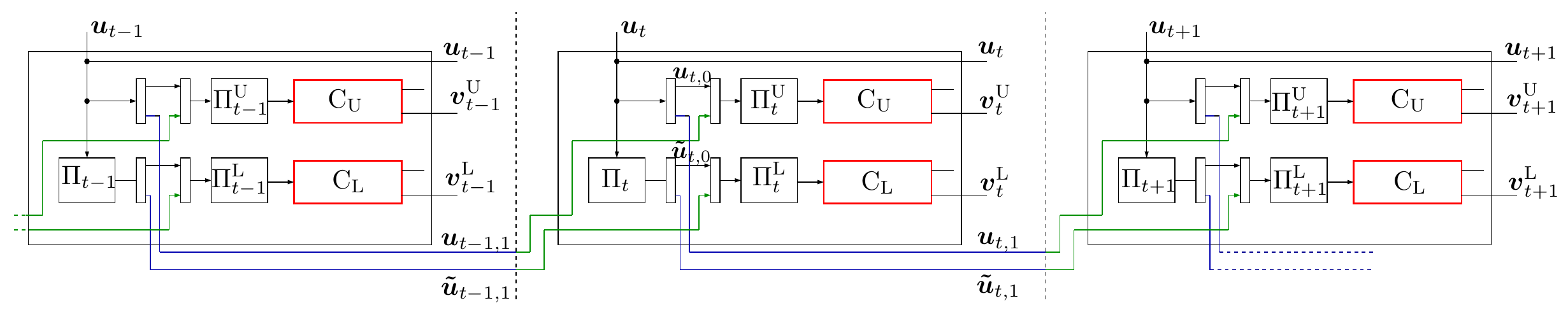}
	\caption{Block diagram of the encoder of a SC-PCC ensemble with $m=1$. }
	\label{CoupledPCC}
\end{figure*}

\begin{figure*}[t]
	\centering
	\includegraphics[width=1\linewidth]{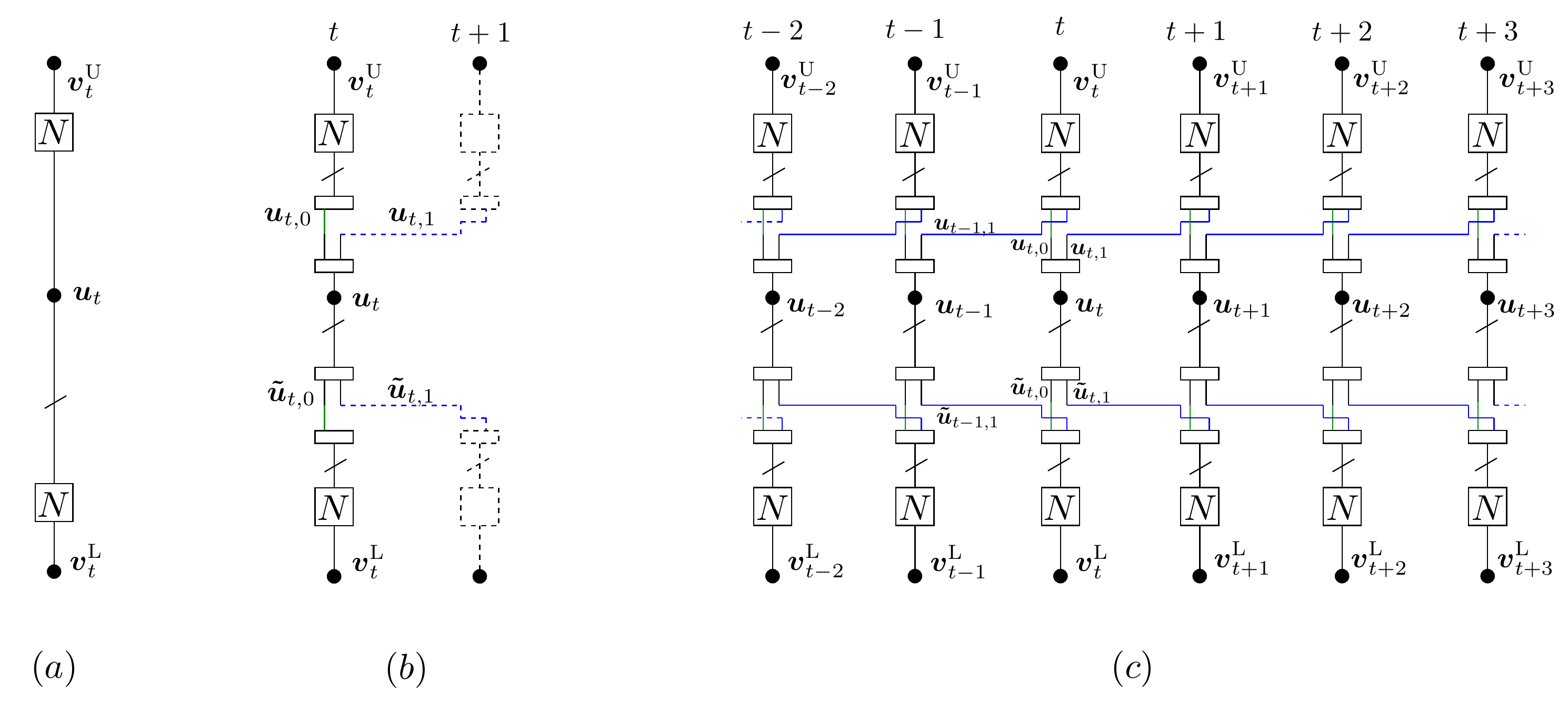}
	\caption{Compact graph representation of (a) PCC, (b) SC-PCC at time instant $t$, (c) SC-PCC.}
	\label{SC-PCC}
\end{figure*}

\subsection{Braided Convolutional Codes}

We consider a rate $R=1/3$ BCC built from two rate-$2/3$ recursive systematic convolutional encoders, referred to as upper and lower encoders. The corresponding trellises are denoted by $\text{T}^{\text{U}}$ and $\text{T}^{\text{L}}$. The compact graph representation of this code is shown in Fig.~\ref{Uncoupled}(d). The parity sequences of the upper and lower encoder are denoted by $\bs{v}^{\text{U}}$ and $\bs{v}^{\text{L}}$, respectively.
To produce the parity sequence $\bs{v}^{\text{U}}$, the information sequence $\bs{u}$ and a reordered copy of  $\bs{v}^{\text{L}}$ are encoded by $\text{T}^{\text{U}}$.
Likewise, a reordered copy of $\bs{u}$ and a reordered copy of $\bs{v}^{\text{U}}$ are encoded by $\text{T}^{\text{L}}$ in order to produce the parity sequence $\bs{v}^{\text{L}}$.
Similarly to PCCs, the transmitted sequence is $\bs{v}=(\bs{u},\bs{v}^{\text{U}},\bs{v}^{\text{L}})$.

\section{Spatially Coupled Turbo-like Codes} 
\label{sec:SCTCs}
In this section, we introduce SC-TCs. We first describe the spatial coupling for both PCCs and SCCs. Then, we generalize the original block-wise BCC ensemble  \cite{ZhangBCC} in order to obtain ensembles with larger coupling memories. 

\subsection{Spatially Coupled Parallel Concatenated Codes}

We consider the spatial coupling of rate-$1/3$ PCCs, described in the previous section.
For simplicity, we first describe the SC-PCC ensemble with coupling memory $m=1$.
Then we show the coupling for higher coupling memories.
The block diagram of the encoder for the SC-PCC ensemble is shown in Fig.~\ref{CoupledPCC}.
In addition, its compact graph representation and the coupling are illustrated in Fig.~\ref{SC-PCC}.


As it is shown in Fig.~\ref{CoupledPCC} and Fig.~\ref{SC-PCC}(a) we denote by $\bs{u}_t$ the information sequence, and by $\bs{v}_t^{\text{U}}$ and $\bs{v}_t^{\text{L}}$ the parity sequence of the upper and lower encoder, respectively, at time $t$.
The code sequence of the PCC at time $t$ is given by the triple $\bs{v}_t = 
(\bs{u}_t,\bs{v}^{\text{U}}_t ,\bs{v}^{\text{L}}_t )$. With reference to Fig.~\ref{CoupledPCC} and Fig.~\ref{SC-PCC}(b),
in order to obtain the coupled sequence, the information sequence, $\bs{u}_t$, is divided into two sequences of equal size, $\bs{u}_{t,0}$ and $\bs{u}_{t,1}$ by a multiplexer. 
Then, the sequence $\bs{u}_{t,0}$ is used as a part of the input to the upper encoder at time $t$ and $\bs{u}_{t,1}$ is used as a part of the input to the upper encoder at time $t+1$.
Likewise, a reordered copy of the information sequence, $\tilde{\bs{u}}_t$, is divided into two sequences $\tilde{\bs{u}}_{t,0}$ and $\tilde{\bs{u}}_{t,1}$.

Therefore, the input to the upper encoder at time $t$ is a reordered copy of $(\bs{u}_{t,0},\bs{u}_{t-1,1})$, and likewise the input to the lower encoder at time $t$ is a reordered copy of $(\tilde{\bs{u}}_{t,0},\tilde{\bs{u}}_{t-1,1})$.
In this ensemble, the coupling memory is $m=1$ as $\bs{u}_t$ is used only at the time instants $t$ and $t+1$.


Finally, an SC-PCC with $m=1$ is obtained by considering a collection of $L$ PCCs at time instants $t=1,\ldots,L$, where $L$ is referred to as the coupling length, and coupling them as described above, see Fig.~\ref{SC-PCC}(c).

\begin{figure}[t]
  \centering
    \includegraphics[width=0.75\linewidth]{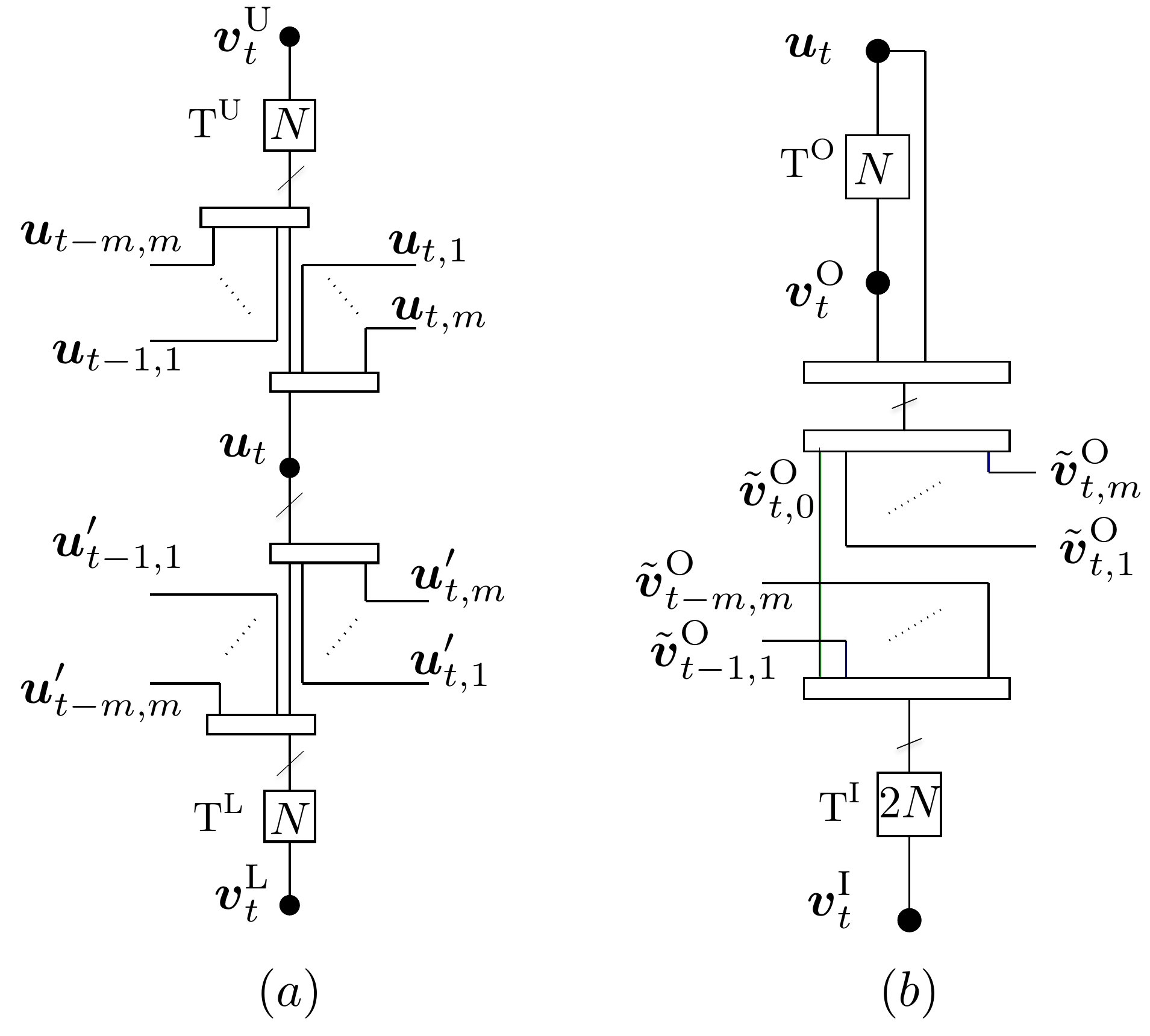}
\caption{Compact graph representation of (a) SC-PCCs, and (b) SC-SCCs of coupling memory $m$ for time instant $t$.}
\label{Coupled}
\end{figure}

An SC-PCC ensemble with coupling memory $m$ is obtained by dividing each of the sequences $\bs{u}_t$ and $\tilde{\bs{u}}_t$ into $m+1$ sequences of equal size and spread these sequences respectively to the input of the upper and the lower encoder at time slots $t$ to $t+m$. The compact graph representation of the SC-PCC with coupling memory $m$ is shown in Fig.~\ref{Coupled}(a) for a given time instant $t$.

The coupling is performed as follows.  Divide the information sequence $\u_t$ into $m+1$ sequences of equal size $N/(m+1)$, denoted by $\bs{u}_{t,j}$, $j=0,\dots,m$. Likewise, divide $\tilde{\bs{u}}_t$, the information sequence $\bs{u}_t$ reordered by a permutation, into $m+1$ sequences of equal size, denoted  by $\tilde{\bs{u}}_{t,j}$, $j=0,\dots,m$. At time $t$, the information sequence at the input of the upper encoder is $(\u_{t,0},\u_{t-1,1},\ldots,\u_{t-m,m})$, properly reordered by a permutation. Likewise, the information sequence at the input of the lower encoder is $(\tilde{\u}_{t,0},\tilde{\u}_{t-1,1},\ldots,\tilde{\u}_{t-m,m})$, reordered by a permutation. 
Using the procedure described above, a coupled chain (a convolutional
structure over time) of $L$ PCCs with coupling memory
$m$ is obtained. 

In order to terminate the encoder of the SC-PCC to the zero state, the information sequences at the end of the chain are chosen in such a way that the code sequences become
$\bs{v}_{t}=\bs{0}$ at time $t=L+1,\dots,L+m$, and $\u_t$ is set to $\bs{0}$ for $t>L$. Analogously to conventional convolutional codes, this results in a rate loss that becomes smaller as $L$ increases.  

\subsection{Spatially Coupled Serially Concatenated Codes}

An SC-SCC is constructed similarly to SC-PCCs. Consider a collection of $L$ SCCs at time instants $t=1,\ldots,L$,
and let $\bs{u}_t$ be the information sequence at time $t$. Also, denote by $\bs{v}_t^{\mathrm{O}}$ and $\bs{v}_t^{\mathrm{I}}$ the parity sequence at the output of the outer and inner encoder, respectively. The information sequence $\bs{u}_t$ and the parity sequence $\bs{v}_t^{\mathrm{O}}$ are multiplexed and reordered into the sequence $\tilde{\bs{v}}_t^{\mathrm{O}}$. The sequence $\tilde{\bs{v}}_t^{\mathrm{O}}$ is divided into $m+1$ sequences of equal length, denoted by $\tilde{\bs{v}}_{t,j}^{\text{O}}$, $j=0,\dots,m$. Then, at time instant $t$, the
sequence at the input of the inner encoder is
$(\tilde{\bs{v}}_{t-j,0}^{\text{O}},\tilde{\bs{v}}_{t-1,1}^{\text{O}}\ldots,\tilde{\bs{v}}_{t-m,m}^{\text{O}})$, properly reordered by a permutation. This sequence is encoded by the inner encoder into $\bs{v}_t^{\mathrm{I}}$. Finally, the code sequence at time $t$ is $\bs{v}=(\bs{u}_t,\bs{v}_t^{\text{O}},\bs{v}_t^{\text{I}})$. Using this construction method, a coupled chain of $L$ SCCs with coupling memory $m$ is obtained. The compact graph representation of SC-SCCs with coupling memory $m$ is shown in Fig.~\ref{Coupled}(b) for time instant $t$. 

In order to terminate the encoder of the SC-SCC, the information sequences at the end of the chain are chosen in such a way that the code sequences become $\bs{v}_{t}=\bs{0}$ at time $t=L+1,\dots,L+m$. A simple and practical way to terminate SC-SCCs is to set $\u_t=\zero$ for $t=L-m+1,\dots,L$. This enforces $\bs{v}_{t}=\bs{0}$ for $t=L+1,\dots,L+m$, since we can assume that $\u_t=\zero$ for $t>L$. Using this termination technique, only the parity sequence $\bs{v}_t^{\text{I}}$ needs to be transmitted at time instants $t=L-m+1,\dots,L$.

\subsection{Braided Convolutional Codes}

\begin{figure}[t]
  \centering
    \includegraphics[width=0.6\linewidth]{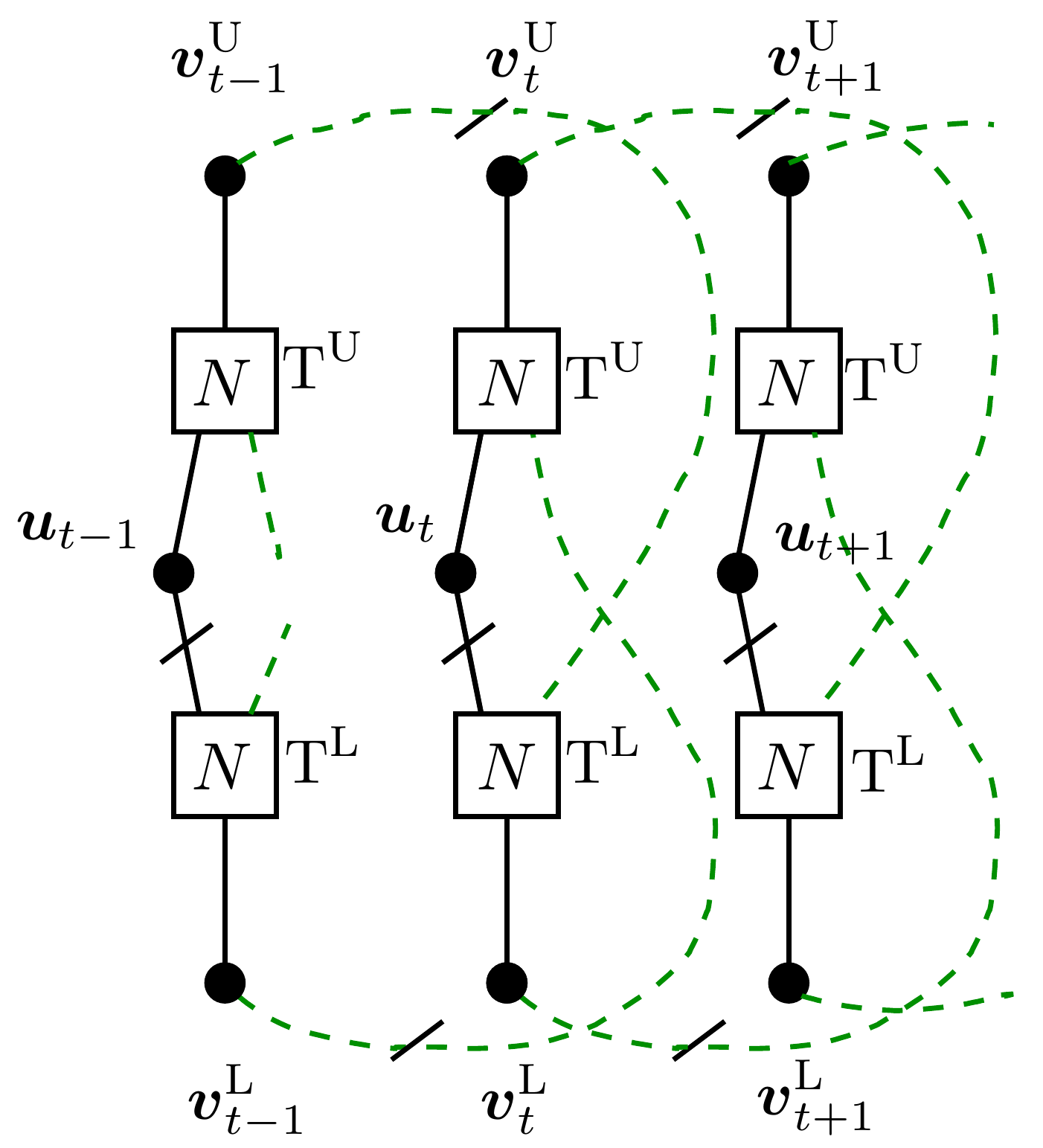}
\caption{Compact graph representation of the original BCCs.}
\label{BCCSC}
\end{figure}

The compact graph representation of the original BCCs is depicted in Fig~\ref{BCCSC}.
As for SC-PCCs, let $\bs{u}_t$, $\bs{v}_t^{\text{U}}$ and $\bs{v}_t^{\text{L}}$ denote the information sequence, the parity sequence at the output of the upper encoder, and the parity sequence at the output of the lower encoder, respectively, at time $t$. 
At time $t$, the information sequence 
 $\bs{u}_t$ and a reordered copy of $\bs{v}_{t-1}^{\text{L}}$ are encoded by the upper encoder to generate the parity sequence $\bs{v}_t^{\text{U}}$.
Likewise, a reordered copy of the information sequence, denoted by $\tilde{\bs{u}}_t$, and a reordered copy of $\bs{v}_{t-1}^{\text{L}}$ are encoded by the lower encoder to produce the parity sequence $\bs{v}_t^{\text{L}}$. The code sequence at time $t$ is $\bs{v}=(\bs{u}_t,\bs{v}_t^{\text{U}},\bs{v}_t^{\text{L}})$.

As it can be seen from Fig~\ref{BCCSC}, the original BCCs are inherently spatially coupled codes\footnote{The uncoupled ensemble, discussed in the previous section, can be defined by tailbiting a coupled chain of length $L=1$.} with coupling memory one. 
In the following, we introduce two extensions of BCCs, referred to as Type-I and Type-II, with increased coupling memory, $m>1$.

\begin{figure}[t]
  \centering
    \includegraphics[width=0.75\linewidth]{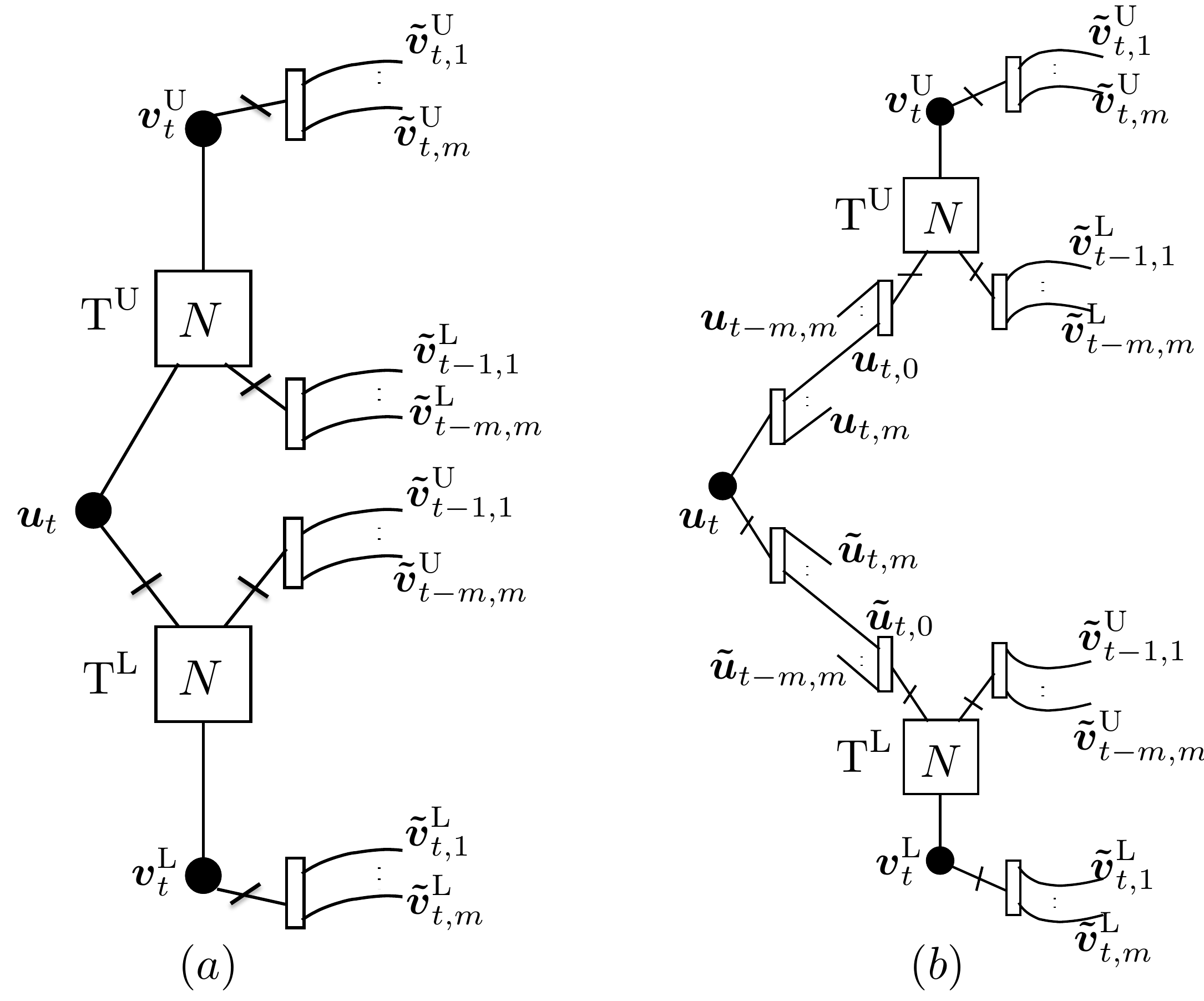}
\caption{Compact graph representation of (a) Type-I BCCs, and (b) Type-II BCCs of coupling memory $m$ at time instant $t$.}
\label{BCCI-II}
\end{figure}

The compact graph of Type-I BCCs is shown in
Fig.~\ref{BCCI-II}(a) for time instant $t$.
The parity sequence $\boldsymbol{v}_t^{\text{U}}$ is randomly divided into $m$ sequences $\boldsymbol{v}_{t,j}^{\text{U}}$, $j=1,\dots,m$, of the same length.
Likewise, the parity sequence $\boldsymbol{v}_t^{\text{L}}$ is randomly divided into $m$ sequences $\boldsymbol{v}_{t,j}^{\text{L}}$, $j=1,\dots,m$. At time $t$, the information sequence $\bs{u}_t$ and the sequence $(\boldsymbol{v}_{t-1,1}^{\text{L}},\boldsymbol{v}_{t-2,2}^{\text{L}},\ldots,\boldsymbol{v}_{t-m,m}^{\text{L}})$, properly reordered, are used as input sequences to the upper encoder to produce the parity sequence $\boldsymbol{v}_t^{\text{U}}$. Likewise, a reordered copy of the information sequence $\bs{u}_t$ and the sequence $(\boldsymbol{v}_{t-1,1}^{\text{U}},\boldsymbol{v}_{t-2,2}^{\text{U}},\ldots,\boldsymbol{v}_{t-m,m}^{\text{U}})$, properly reordered, are encoded by the lower encoder to produce the parity sequence $\boldsymbol{v}_t^{\text{L}}$. 

The compact graph of Type-II BCCs is shown in
Fig.~\ref{BCCI-II}(b) for time instant $t$. Contrary to Type-I BCCs, in addition to the coupling of parity bits, for Type-II BCCs information bits are also coupled. At time $t$, divide the information sequence $\u_t $ into $m+1$ sequences $\boldsymbol{u}_{t,j}$, $j=0,\dots,m$ of equal length. Furthermore, divide the reordered copy of the information sequence, $\tilde{\bs{u}}_t$, into $m+1$ sequences $\tilde{\boldsymbol{u}}_{t,j}$, $j=0,\dots,m$. The first input of the upper and lower encoders are now the sequences $(\bs{u}_{t-0,0},\bs{u}_{t-1,1},\ldots,\bs{u}_{t-m,m})$ and $(\tilde{\bs{u}}_{t-0,0},\tilde{\bs{u}}_{t-1,1},\ldots,\tilde{\bs{u}}_{t-m,m})$, respectively, properly reordered.


\section{Density Evolution Analysis for SC-TCs over the Binary Erasure Channel}\label{sec:DE}

In this section we derive the exact DE for SC-TCs. For the three considered code ensembles, we first derive the DE equations for the uncoupled ensembles and then extend them to the coupled ones.
\subsection{Density Evolution Equations and Decoding Thresholds} 
For transmission over the BEC, it is possible to analyze the asymptotic behavior of TCs and SC-TCs by tracking the evolution of the erasure probability with the number of decoding iterations.
This evolution can be formalized in a compact way as a set of equations called DE equations. For the BEC, it is possible to derive a exact DE equations for TCs and SC-TCs.
By use of these equations, the BP decoding threshold can be computed.
The BP threshold is the largest channel erasure probability $\varepsilon$ for which the erasure probability at the output of the BP decoder converges to zero as the block length and number of iterations grow to infinity.

It is also possible to compute the threshold of the MAP decoder, $\varepsilon_{\text{MAP}}$, by the use of the area theorem \cite{Measson_Turbo}.
According to the area theorem, the MAP threshold\footnote{The threshold given by the area theorem is actually an upper bound on the MAP threshold. However, the numerical results show that the thresholds of the coupled ensembles converge to this upper bound. This indicates that the upper bound on the MAP threshold is a tight bound.} 
can be obtained from the following equation,
\[
\int_{\varepsilon_{\text{MAP}}}^1\bar{p}_{\text{extr}}(\varepsilon)d\varepsilon=R \ ,
\]
where $R$ is the rate of the code and $\bar{p}_{\text{extr}}(\varepsilon)$ is the average extrinsic erasure probability for all transmitted bits.

\subsection{Parallel Concatenated Codes}
\subsubsection{Uncoupled}
Consider the compact graph of a PCC in Fig.~\ref{Uncoupled}(b). 
Let $p_{\text{U},\text{s}}^{(i)}$ and $p_{\text{U},\text{p}}^{(i)}$ denote 
the average extrinsic erasure probability from factor node $T^{\text{U}}$ to $\bs{u}$ and $\bs{v}^{\text{U}}$, respectively, in the $i$th iteration.\footnote{With some abuse of language, we sometimes refer to a variable node representing a sequence (e.g., $\u$) as the sequence itself ($\u$ in this case).} Likewise, denote by $p_{\text{L},\text{s}}^{(i)}$ and $p_{\text{L},\text{p}}^{(i)}$ the extrinsic erasure probabilities from $T^{\text{L}}$ to $\bs{u}$ and $\bs{v}^{\text{L}}$, respectively. It is easy to see that the erasure probability from $\bs{u}_t$ and $\bs{v}_t^{\text{U}}$ to $T^{\text{U}}$ is $\varepsilon \cdot p_{\text{L},\text{s}}^{(i-1)}$ and $\varepsilon$, respectively. Therefore, the DE updates for $T^{\text{U}}$ can be written as
\begin{align}
\label{DEPCC1}
p_{\text{U},\text{s}}^{(i)}=f_{\text{U,s}}\left(
q_{\text{L}}^{(i)},\varepsilon\right),\\
\label{DEPCC2}
p_{\text{U},\text{p}}^{(i)}=f_{\text{U,p}}\left(
q_{\text{L}}^{(i)},\varepsilon\right),
\end{align}
where
\begin{equation}
\label{DEPCC3}
q_{\text{L}}^{(i)}=\varepsilon \cdot p_{\text{L},\text{s}}^{(i-1)},
\end{equation}
and $f_{\text{U,s}}$ and $f_{\text{U,p}}$ denote the transfer function of $T^{\text{U}}$ for the systematic and parity bits, respectively. 

Similarly, the DE update for $T^{\text{L}}$ can be written as
\begin{align}
\label{DEPCC4}
p_{\text{L},\text{s}}^{(i)}=f_{\text{L,s}}\left(
q_{\text{U}}^{(i)},\varepsilon\right),\\ \label{DEPCC5}
p_{\text{L},\text{p}}^{(i)}=f_{\text{L,p}}\left(
q_{\text{U}}^{(i)},\varepsilon\right),
\end{align}
where
\begin{equation}
\label{DEPCC6}
q_{\text{U}}^{(i)}=\varepsilon \cdot p_{\text{U},\text{s}}^{(i-1)},
\end{equation}
and $f_{\text{L,s}}$ and $f_{\text{L,p}}$ are the transfer functions of  $T^{\text{L}}$ for the systematic and parity bits, respectively.
\subsubsection{Coupled}
Consider the compact graph of a SC-PCC ensemble in Fig.~\ref{Coupled}(a).
The variable node $\bs{u}_t$ is connected to factor nodes $T^{\text{U}}_{t'}$ and $T^{\text{L}}_{t'}$, at time instants $t'=t,\ldots,t+m$. We denote by $p_{\text{U},\text{s}}^{(i,t')}$ and $p_{\text{U},\text{p}}^{(i,t')}$  
the average extrinsic erasure probability from factor node $T^{\text{U}}_{t'}$ at time instant $t'$ to $\bs{u}$ and $\bs{v}^{\text{U}}$, respectively, computed in the $i$th iteration. We also denote by $\bar{q}^{(i-1,t)}_{\text{U}}$ the input erasure probability to variable node $\bs{u}_t$ in the $i$th iteration, received from its neighbors $T^{\text{U}}_{t'}$. It can be written as
\begin{equation}\label{DESCPCC1}
\bar{q}_{\text{U}}^{(i-1,t)}=\frac{1}{m+1} \sum_{j=0}^{m} p_{\text{U},\text{s}}^{(i-1,t+j)}.
\end{equation}

Similarly, the average erasure probability from factor nodes $T^{\text{L}}_{t'}$, $t'=t,\ldots,t+m$, to $\bs{u}_t$, denoted by $\bar{q}_{\text{L}}^{(i-1,t)}$, can be written as
\begin{equation}\label{DESCPCC2}
\bar{q}_{\text{L}}^{(i-1,t)}=\frac{1}{m+1} \sum_{j=0}^{m} p_{\text{L},\text{s}}^{(i-1,t+j)}.
\end{equation} 

The erasure probabilities from variable node $\bs{u}_t$ to its neighbors $T^{\text{U}}_{t'}$ and  $T^{\text{L}}_{t'}$ are
 $\varepsilon \cdot \bar{q}^{(i-1,t)}_{\text{L}}$ and $\varepsilon \cdot \bar{q}^{(i-1,t)}_{\text{U}}$, respectively.

On the other hand, $T^{\text{U}}_t$ at time $t$ is connected to the set of $\bs{u}_{t'}$s for $t'=t-m, \dots, t$. 
The erasure probability to $T^{\text{U}}_t$ from this set, denoted by $q_{\text{L}}^{(i,t)}$, is given by
\begin{align}
\label{DESCPCC3}
q_{\text{L}}^{(i,t)}&=\varepsilon \cdot \frac{1}{m+1} \sum_{k=0}^{m}
\bar{q}_{\text{L}}^{(i-1,t-k)}\nonumber\\
&=\varepsilon \cdot \frac{1}{(m+1)^2} \sum_{k=0}^{m}\sum_{j=0}^{m} p_{\text{L},\text{s}}^{(i-1,t+j-k)}.
\end{align}

Thus, the DE updates of $T^{\text{U}}_t$ are
\begin{align}
\label{DESCPCC4}
p_{\text{U},\text{s}}^{(i,t)}=f_{\text{U,s}}\left(
q_{\text{L}}^{(i,t)},\varepsilon\right),\\
\label{DESCPCC5}
p_{\text{U},\text{p}}^{(i,t)}=f_{\text{U,p}}\left(
q_{\text{L}}^{(i,t)},\varepsilon\right).
\end{align}

Similarly, the input erasure probability to $T^{\text{L}}_t$ from the set of connected $\bs{u}_{t'}$s at time instants $t'=t-m, \dots, t$, is
\begin{align}
\label{DESCPCC6}
&q_{\text{U}}^{(i,t)}=\varepsilon \cdot \frac{1}{m+1} \sum_{k=0}^{m}
\bar{q}_{\text{U}}^{(i-1,t-k)} \nonumber\\
&=\varepsilon \cdot \frac{1}{(m+1)^2} \sum_{k=0}^{m}\sum_{j=0}^{m} p_{\text{U},\text{s}}^{(i-1,t+j-k)},
\end{align} 
and the DE updates of $T^{\text{L}}_t$ are
\begin{align}
\label{DESCPCC7}
p_{\text{L},\text{s}}^{(i,t)}=f_{\text{L,s}}\left(
q_{\text{U}}^{(i,t)},\varepsilon\right),\\
\label{DESCPCC8}
p_{\text{L},\text{p}}^{(i,t)}=f_{\text{L,p}}\left(
q_{\text{U}}^{(i,t)},\varepsilon\right).
\end{align}

Finally the a-posteriori erasure probability on $\bs{u}_t$ at time $t$ and iteration $i$ is
\begin{equation}
p_a^{(i,t)}=\varepsilon \cdot \bar{q}_{\text{U}}^{(i,t)} \cdot \bar{q}_{\text{L}}^{(i,t)}.
\end{equation}
 DE is performed by tracking the evolution of the a-posteriori erasure probability
 with the number of iterations.
\subsection{Serially Concatenated Codes}
\subsubsection{Uncoupled}
Consider the compact graph of the SCC ensemble in Fig.~\ref{Uncoupled}(c).
Let $p_{\text{O},\text{s}}^{(i)}$ and $p_{\text{O},\text{p}}^{(i)}$ denote the erasure probability from $T^{\text{O}}$ to $\bs{u}$ and $\bs{v}^{\text{O}}$, respectively, computed in the $i$th iteration.
Likewise, $p_{\text{I},\text{s}}^{(i)}$ and $p_{\text{I},\text{p}}^{(i)}$ denote the extrinsic erasure probability from $T^{\text{I}}$ to  $\tilde{\bs{v}}^{\text{O}}=(\bs{u},\bs{v}^{\text{O}})$ and $\bs{v}^{\text{I}}$. 

Both $\bs{u}$ and $\bs{v}^{\text{O}}$ receive the same erasure probability, $p_{\text{I},\text{s}}^{(i-1)}$, from $T^{\text{I}}$. Therefore, the erasure probabilities that $T^{\text{O}}$ receives from these two variable nodes are equal and given by
\begin{align}
q_{\text{I}}^{(i)}=\varepsilon \cdot p_{\text{I},\text{s}}^{(i-1)}.\label{DESCC1}
\end{align}
The DE equations for $T^{\text{O}}$ can then be written as
\begin{align}
\label{DESCC2}
p_{\text{O},\text{s}}^{(i)}=f_{\text{O,s}}\left(
q_{\text{I}}^{(i)},q_{\text{I}}^{(i)}\right),\\
\label{DESCC3}
p_{\text{O},\text{p}}^{(i)}=f_{\text{O,p}}\left(
q_{\text{I}}^{(i)},q_{\text{I}}^{(i)}\right),
\end{align}
where $f_{\text{O,s}}$ and $f_{\text{O,p}} $ are the transfer functions of $T^{\text{O}}$ for the systematic and parity bits, respectively. 

 The erasure probability that $T^{\text{I}}$ receives from $\tilde{\bs{v}}^{\text{O}}=(\bs{u},\bs{v}^{\text{O}})$ is the average of the erasure probabilities from  $\bs{u}$ and $\bs{v}^{\text{O}}$,
\begin{equation}
\label{DESCC4}
q_{\text{O}}^{(i)}=\varepsilon \cdot\frac{p_{\text{O},\text{s}}^{(i)}+p_{\text{O},\text{p}}^{(i)}}{2}.
\end{equation}
On the other hand, the erasure probability to $T^{\text{I}}$ from $\bs{v}^{\text{I}}$ is $\varepsilon$. Therefore, the DE equations for $T^{\text{I}}$ can be written as
\begin{align}
\label{DESCC5}
p_{\text{I},\text{s}}^{(i)}=f_{\text{I,s}}\left(
q_{\text{O}}^{(i)},\varepsilon\right),
\\
\label{DESCC6}
p_{\text{I},\text{p}}^{(i)}=f_{\text{I,p}}\left(
q_{\text{O}}^{(i)},\varepsilon\right).
\end{align}
\subsubsection{Coupled}
Consider the compact graph representation of SC-SCCs in Fig.~\ref{Coupled}(b). 
Variable nodes $\bs{u}_t$ and $\bs{v}_t^{\text{O}}$ are connected to factor nodes $T^{\text{I}}_{t'}$ at time instants $t'=t,\ldots,t+m$. 
The input erasure probability to variable nodes $\bs{u}_t$ and $\bs{v}_t^{\text{O}}$ from these factor nodes, denoted by $\bar{q}_{\text{I}}^{(i-1,t)}$, is the same for both $\bs{u}_t$ and $\bs{v}_t^{\text{O}}$ and is obtained as the average of the erasure probabilities from each of the factor nodes $T^{\text{I}}_{t'}$,
\begin{equation}
\label{DESCSCC1}
\bar{q}_{\text{I}}^{(i-1,t)}=\frac{1}{m+1}\sum_{j=0}^{m}p_{\text{I},\text{s}}^{(i-1,t+j)} \ .
\end{equation}
The erasure probability to $T^{\text{O}}_{t}$ from $\bs{u}_t$ and $\bs{v}_t^{\text{O}}$ is
\begin{align}
\label{DESCSCC2}
q_{\text{I}}^{(i,t)}=\varepsilon \cdot \bar{q}_{\text{I}}^{(i-1,t)}= \frac{\varepsilon}{m+1}\sum_{j=0}^{m}p_{\text{I},\text{s}}^{(i-1,t+j)} \ .
\end{align}
Thus, the DE updates of $T^{\text{O}}_t$ are
\begin{align}
\label{DESCSCC3}
p_{\text{O},\text{s}}^{(i,t)}=f_{\text{O,s}}\left(
q_{\text{I}}^{(i,t)},q_{\text{I}}^{(i,t)}\right) \ ,\\
\label{DESCSCC4}
p_{\text{O},\text{p}}^{(i,t)}=f_{\text{O,p}}\left(
q_{\text{I}}^{(i,t)},q_{\text{I}}^{(i,t)}\right) \ . 
\end{align}
At time $t$, $T^{\text{I}}_t$ is connected to a set of $\tilde{\bs{v}}_{t'}^{\text{O}}$s at time instants $t'=t-m,\ldots,t$. 
The erasure probability that $T^{\text{I}}_t$ receives from this set is the average of the erasure probabilities of all $\bs{u}_{t'}$s and $\bs{v}_{t'}^{\text{O}}$s at times $t'=t-m\ldots,t$. This erasure probability can be written as
 \begin{align}
\label{DESCSCC5}
q_{\text{O}}^{(i,t)}=\frac{\varepsilon}{m+1}\sum_{k=0}^{m}\frac{p_{\text{O},\text{s}}^{(i,t-k)}+p_{\text{O},\text{p}}^{(i,t-k)}}{2} \ .
\end{align}
Hence, the DE updates for the inner encoder are given by
\begin{align}
\label{DESCSCC6}
p_{\text{I},\text{s}}^{(i,t)}=f_{\text{I,s}}\left(
q_{\text{O}}^{(i,t)},\varepsilon\right) \ ,
\\
\label{DESCSCC7}
p_{\text{I},\text{p}}^{(i,t)}=f_{\text{I,p}}\left(
q_{\text{O}}^{(i,t)},\varepsilon\right) \ .
\end{align}
Finally, the a-posteriori erasure probability on information bits at time $t$ and iteration $i$ is
\begin{equation}
p_{a}^{(i,t)}=\varepsilon \cdot p_{\text{O},\text{s}}^{(i,t)}\cdot \bar{q}_{\text{I}}^{(i,t)} \ .
\end{equation}
\subsection{Braided Convolutional Codes}
\subsubsection{Uncoupled}
Consider the compact graph of uncoupled BCCs in Fig.~\ref{Uncoupled}(c). These can be obtained by tailbiting BCCs, as shown in Fig.~\ref{BCCSC}, with coupling length $L=1$.
Let $p_{\text{U},k}^{(i)}$ and $p_{\text{L},k}^{(i)}$ denote the erasure probabilities of messages from  $T^{\text{U}}$ and $T^{\text{L}}$ through their $k$th connected edge, $k=1,2,3$, respectively.
The erasure probability of messages that $T^{\text{U}}$ receives through its edges are
\begin{align}
q_{\text{L},1}^{(i)}=\varepsilon \cdot p_{\text{L},1}^{(i-1)} \ ,\label{DEBCC1}\\
q_{\text{L},2}^{(i)}=\varepsilon \cdot p_{\text{L},3}^{(i-1)} \ ,\label{DEBCC2}\\
q_{\text{L},3}^{(i)}=\varepsilon \cdot
p_{\text{L},2}^{(i-1)} \ .\label{DEBCC3}
\end{align}
The exact DE equations of $T^{\text{U}}$ can be written as
\begin{align}
p_{\text{U},1}^{(i)}=&f_{\text{U},1}\left(q_{\text{L},1}^{(i)} ,q_{\text{L},2}^{(i)},q_{\text{L},3}^{(i)}\right) \label{DEBCC4}\ ,\\
p_{\text{U},2}^{(i)}=&f_{\text{U},2}\left(q_{\text{L},1}^{(i)} ,q_{\text{L},2}^{(i)},q_{\text{L},3}^{(i)}\right) \label{DEBCC5}\ ,\\
p_{\text{U},3}^{(i)}=&f_{\text{U},3}\left(q_{\text{L},1}^{(i)} ,q_{\text{L},2}^{(i)},q_{\text{L},3}^{(i)}\right)  \label{DEBCC6}\ , 
\end{align}
where $f_{\text{U},k}$ denotes the transfer function of $T^{\text{U}}$ for its $k$th connected edge.
Similarly, the DE equations for $T^{\text{L}}$ can be written by swapping indexes $\text{U}$ and $\text{L}$ in \eqref{DEBCC1}--\eqref{DEBCC6}.
\subsubsection{Coupled}
Consider the compact graph representation of Type-I BCCs in Fig.~\ref{BCCI-II}(a).
As in the uncoupled case, the DE updates of factor nodes $T^{\text{U}}_t$ and $T^{\text{L}}_t$ are similar due to the symmetric structure of the coupled construction. Therefore, for simplicity, we only describe the DE equations of $T^{\text{U}}_t$ and the equations for $T^{\text{L}}_t$ are obtained by swapping indexes $\text{U}$ and $\text{L}$ in the equations.

The first edge of $T^{\text{U}}_t$ is connected to $\bs{u}_t$. Thus, the erasure probability that $T^{\text{U}}_t$ receives through this edge is
\begin{equation}
\label{eq:T1BCCDE1}
q_{\text{L},1}^{(i,t)}=\varepsilon \cdot  p_{\text{L},1}^{(i-1,t)}.
\end{equation}
The second edge of $T^{\text{U}}_t$ is connected to variable nodes $\bs{v}_{t'}^{\text{L}}$ at time instants $t'=t-m,\ldots,t-1$.
 The erasure probability that $T^{\text{U}}_t$ receives through its second edge is therefore the average of the erasure probabilities from the variable nodes $\bs{v}_{t'}^{\text{L}}$ that are connected to this edge. 
This erasure probability can be written as
\begin{equation}
\label{eq:T1BCCDE2}
q_{\text{L},2}^{(i,t)}=\frac{\varepsilon}{m}\sum_{j=1}^{m}  p_{\text{L},3}^{(i-1,t-j)}.
\end{equation}
The third edge of $T^{\text{U}}_t$ is connected to $\bs{v}_t^{\text{U}}$, which is in turn connected to the second edges of factor nodes $T^{\text{L}}_{t'}$ at time instants $t'=t+1,\ldots,t+m$. 
The erasure probability that $\bs{v}_t^{\text{U}}$ receives from the set of connected nodes $T^{\text{L}}_{t'}$ is the average of erasure probabilities from these nodes through their second edges.
The erasure probability from $\bs{v}_t^{\text{U}}$ to $T^{\text{U}}_t$ is
\begin{equation}
\label{eq:T1BCCDE3}
q_{\text{L},3}^{(i,t)}=\frac{\varepsilon}{m}\sum_{j=1}^{m}  p_{\text{L},2}^{(i-1,t+j)}.
\end{equation}
The DE equations of $T^{\text{U}}_t$ can then be written as\footnote{The DE equations of the original BCCs are obtained by setting $m=1$ in the DE equations of Type-I BCCs.}
\begin{align}
\label{eq:T1BCCDE4}
p_{\text{U},1}^{(i,t)}=&f_{\text{U},1}\left(q_{\text{L},1}^{(i,t)},q_{\text{L},2}^{(i,t)},q_{\text{L},3}^{(i,t)}\right),\\ \label{eq:T1BCCDE5}
p_{\text{U},2}^{(i,t)}=&f_{\text{U},2}\left(q_{\text{L},1}^{(i,t)},q_{\text{L},2}^{(i,t)},q_{\text{L},3}^{(i,t)}\right),\\ \label{eq:T1BCCDE6}
p_{\text{U},3}^{(i,t)}=&f_{\text{U},3}\left(q_{\text{L},1}^{(i,t)},q_{\text{L},2}^{(i,t)},q_{\text{L},3}^{(i,t)}\right).
\end{align} 
The a-posteriori erasure probability on $\bs{u}_t$ at time $t$ and iteration $i$ for Type-I BCCs is
\begin{equation}
p_a^{(i,t)}=\varepsilon \cdot  p_{\text{U},1}^{(i,t)} \cdot p_{\text{L},1}^{(i,t)}.
\end{equation}

As we discussed in the previous section, the difference between Type-I and Type-II BCCs is that $\bs{u}_t$ is also coupled in the latter. Variable node
$\bs{u}_t$ in Type-II BCCs is connected to a set of factor nodes $T^{\text{U}}_{t'}$ and $T^{\text{L}}_{t'}$ at time instants $t'=t,\ldots,t+m$.
The DE equations of Type-II BCCs are identical to those of Type-I BCCs except for equation \eqref{eq:T1BCCDE1}. Denote by $\bar{q}_{\text{L},1}^{(i-1,t)}$ the input erasure probability to $\bs{u}_t$ from the connected factor nodes $T^{\text{L}}_{t'}$ in the $i$th iteration.
According to Fig.~\ref{BCCI-II}(b), $\bar{q}_{\text{L},1}^{(i-1,t)}$ is the average of erasure probabilities
from $T^{\text{L}}_{t'}$ at time instants $t'=t,\ldots,t+m$, 
\begin{equation}
\bar{q}_{\text{L},1}^{(i-1,t)}=\frac{1}{m+1} \sum_{j=0}^{m} p_{\text{L},1}^{(i-1,t+j)}.
\end{equation}
Factor node $T^{\text{U}}_{t}$ is connected to variable nodes $\bs{u}_{t'}$ at time instants $t'=t-m,\ldots,t$. 
The incoming erasure probability to $T^{\text{U}}_t$ through its first edge, denoted by $q_{\text{L},1}^{(i,t)}$, is therefore the average of the erasure probabilities from $\bs{u}_{t'}$ at times $t'=t-m,\ldots,t$, 
 \begin{align}
&q_{\text{L},1}^{(i,t)}=\varepsilon \cdot \frac{1}{m+1} \sum_{k=0}^{m}
\bar{q}_{\text{L},1}^{(i-1,t-k)}\\
&=\varepsilon \cdot \frac{1}{(m+1)^2} \sum_{k=0}^{m}\sum_{j=0}^{m} p_{\text{L},1}^{(i-1,t+j-k)}.\nonumber
\end{align}
Finally, the a-posteriori erasure probability on $\bs{u}_t$ at time $t$ and iteration $i$ for Type-II BCCs is
\begin{equation}
p_a^{(i,t)}=\varepsilon \cdot \bar{q}_{\text{U}}^{(i,t)} \cdot \bar{q}_{\text{L}}^{(i,t)}.
\end{equation}


\section{Rate-compatible SC-TCs via Random Puncturing}\label{sec:RandomP}

Higher rate codes can be obtained by applying puncturing. For analysis purposes, we consider random puncturing.  Random puncturing has been considered, e.g., for LDPC codes in \cite{PishroNik, LDPCPuncture} and for turbo-like codes in \cite{AGiAa,Kol12}. In \cite{LDPCPuncture}, the authors introduced a parameter called $\theta$ which allows comparing the strengths of the codes with different rates. In this section, we consider the construction of rate-compatible SC-TCs by means of random puncturing.

We denote by $\rho\in[0,1]$ the fraction of surviving bits after puncturing, referred to as the permeability rate. Consider that a code sequence $\boldsymbol{v}$ is randomly punctured with permeability rate $\rho$ and transmitted over a BEC with erasure probability $\varepsilon$, BEC$(\varepsilon)$. For the BEC, applying puncturing  is equivalent to transmitting $\bs{v}$ over a BEC with erasure probability $\varepsilon_{\rho}=1-(1-\varepsilon)\rho$, resulting from the concatenation of two BECs, BEC$(\varepsilon)$ and BEC$(\varepsilon_\rho)$. The DE equations of SC-TCs in the previous section can then be easily modified to account for random puncturing.

For SC-PCCs, we consider puncturing of parity bits only, i.e., the overall code is systematic. The rate of the punctured code (without considering termination of the coupled chain) is
$R=\frac{1}{1+2\rho}$. The DE equations of punctured SC-PCCs are obtained
by substituting $\varepsilon\leftarrow \varepsilon_{\rho}$ in \eqref{DESCPCC4}, \eqref{DESCPCC5}, \eqref{DESCPCC7} and \eqref{DESCPCC8}. 

For punctured SC-SCCs, we consider the coupling of the punctured SCCs proposed in \cite{AGiAa,AGiAb}\footnote{ In contrast to standard SCCs, characterized by a rate-$1$ inner code and for which to achieve higher rates the outer code is heavily punctured, the SCCs proposed in \cite{AGiAa,AGiAb} achieve higher rates by moving the puncturing of the outer code to the inner code, which is punctured beyond the unitary rate. This allows to preserve the interleaving gain for high rates and yields a larger minimum distance, which results in codes that significantly outperform standard SCCs, especially for high rates. Furthermore, the SCCs in \cite{AGiAa,AGiAb} yield better MAP thresholds than standard SCCs.}, where $\rho_0$ and $\rho_1$ are the permeability rates of the systematic and
parity bits, respectively, of the outer code (see
\cite[Fig.~1]{AGiAb}), and $\rho_2$ is the permeability rate of the
parity bits of the inner code. The code rate of the 
punctured SC-SCC is
$R=\frac{1}{\rho_0+\rho_1+2\rho_2}$ (neglecting the rate loss due to termination). The DE for punctured SC-SCCs is
obtained by substituting $\varepsilon\leftarrow\varepsilon_{\rho_2}$
in \eqref{DESCSCC6} and \eqref{DESCSCC7}, and modifying \eqref{DESCSCC5} to
\begin{align*}
q_{\text{O}}^{(i,t)}=\frac{1}{m+1}\sum_{k=0}^{m}\frac{\varepsilon
  \cdot p_{\text{O},\text{s}}^{(i,t-k)}
+\varepsilon_{\rho_1} \cdot p_{\text{O},\text{p}}^{(i,t-k)}}{2}
\end{align*}
and \eqref{DESCSCC3}, \eqref{DESCSCC4} to
\begin{align}
p_{\text{O},\text{s}}^{(i,t)}=f_{\text{O,s}}\left(
q_{\text{I}}^{(i,t)},\tilde{q}_{\text{I}}^{(i,t)}\right),\\
p_{\text{O},\text{p}}^{(i)}=f_{\text{O,p}}\left(
q_{\text{I}}^{(i,t)},\tilde{q}_{\text{I}}^{(i,t)}\right),
\end{align}
where $q_{\text{I}}^{(i,t)}$ is given in \eqref{DESCSCC2} and
\begin{equation}
\tilde{q}_{\text{I}}^{(i,t)}=\frac{\varepsilon_{\rho_1}}{m+1}\sum_{j=0}^{m}p_{\text{I},\text{s}}^{(i-1,t+j)}.
\end{equation}

For both Type-I and Type-II BCCs, similarly to SC-PCCs, we consider only puncturing of parity bits with permeability rate $\rho$.
The DE equations of punctured SC-BCCs are obtained by substituting $\varepsilon\leftarrow \varepsilon_{\rho}$ in \eqref{eq:T1BCCDE2} and \eqref{eq:T1BCCDE3} and the corresponding equations for $q_{\text{U},2}^{(i,t)}$ and $q_{\text{U},3}^{(i,t)}$.

\section{Numerical Results}\label{Sec6}

In Table~\ref{Tab:BPThresholds}, we give DE results for the SC-TC ensembles, and their uncoupled ensembles for rate  $R=1/2$.
In particular, we consider SC-PCC and SC-SCC ensembles with identical 4-state and 8-state component encoders with generator matrix ${\bf{G}}=(1,5/7)$ and ${\bf{G}}=(1,11/13)$, respectively, in octal notation.
For the BCC ensemble, we consider two identical 4-state component encoders and generator matrix
\begin{equation}\label{eqG}
\bs{G}_1 (D)= \left( \begin{array}{ccc}1&0&1/7\\0&1&5/7\end{array}\right) \ .
\end{equation}
The BP thresholds ($\varepsilon_{\text{BP}}$) and MAP thresholds ($\varepsilon_{\text{MAP}}$) for the uncoupled ensembles are reported in Table~\ref{Tab:BPThresholds}.
The MAP threshold is obtained using the area theorem \cite{AshikhminEXIT,Measson2009}.
We also give the BP thresholds of SC-TCs for coupling memory $m=1$, denoted by $\varepsilon_{\text{SC}}^{1}$.
\begin{table}[t]
	\caption{Thresholds for rate-$1/2$ TCs, and SC-TCs}
	\begin{center}
		\begin{tabular}{lcccc}
			\hline
			Ensemble&states&$\varepsilon_{\text{BP}}$ & $\varepsilon_{\text{MAP}}$  &$\epsSCone$ \\
			\hline
			$\CPCCa$/$\CPCCSa$&4&0.4606&0.4689&0.4689\\[0.5mm]
			$\CSCCa$/$\CSCCSa$&4&0.3594&0.4981& 0.4708\\[0.5mm]
			$\CPCCa$/$\CPCCSa$&8&0.4651&0.4863&0.4862\\[0.5mm]
			$\CSCCa$/$\CSCCSa$&8&0.3120&0.4993& 0.4507\\
			Type-I $\CBCC$&4&0.3013 &0.4993&0.4932\\[0.5mm]
			Type-II $\CBCC$&4& 0.3013&0.4993& 0.4988\\[0.5mm]
			\hline
			
		\end{tabular} 
	\end{center}
	\label{Tab:BPThresholds} 
\end{table}

As expected, PCC ensembles yield better BP thresholds than SCC ensembles. However, SCCs have better MAP threshold. The BP decoder works poorly for uncoupled BCCs and the BP thresholds are worse than those of PCCs and SCCs. On the other hand, the MAP thresholds of BCCs are better than those of both PCCs and SCCs. By applying coupling, the BP threshold improves and for $m=1$, the Type-II BCC ensemble has the best coupling threshold.

Table \ref{Tab:BPThresholdsSCC} shows the thresholds of TCs and SC-TCs for several rates. In the table, for the ensembles $\CPCCa$/$\CPCCSa$, $\rho_2$ is the permeability rate of the parity bits of the upper encoder and the lower encoder. For example, $\rho_2=0.5$ means that half of the bits of $\bs{v}^{\text{U}}$ and $\bs{v}^{\text{L}}$ are punctured (thus, the resulting code rate is $R=1/2$). Note that $\rho_2$ corresponds to permeability $\rho$ defined in Section~\ref{sec:RandomP}. Here, we use $\rho_2$ instead to unify notation with that of SCCs. For the ensembles $\CSCCa$/$\CSCCSa$ (based on the SCCs introduced in \cite{AGiAa,AGiAb}), for a given code rate $R$ the puncturing rates $\rho_0$, $\rho_1$ and $\rho_2$ (see Section~\ref{sec:RandomP}) may be optimized. In this paper, we consider $\rho_0=1$, i.e., the overall code is systematic, and we optimize $\rho_1$ and $\rho_2$ such that the MAP threshold of the (uncoupled) SCC is maximized.\footnote{We remark that nonsystematic codes, i.e., $\rho_0<1$, lead to better MAP thresholds. In this case, the optimum is to puncture last the parity bits of the inner encoder, i.e., for $R<1/2$ $\rho_2=1$ and for $R\ge 1/2$ $\rho_0=0$, $\rho_1=0$ and $\rho_2=1/2R$.} Note that, if $\rho_0=1$, for a given $R$ the optimization simplifies to the optimization of a single parameter, say $\rho_2$, since $\rho_1$ and $\rho_2$ are related by $\rho_1=\frac{1}{R}-1-2\rho_2$.\footnote{Alternatively, one may optimize $\rho_1$ and $\rho_2$ such that the BP threshold of the SC-SCC is optimized for a given coupling memory $m$.} Rate-compatibility can be guaranteed by choosing $\rho_1$ and $\rho_2$ to be decreasing functions of $R$. In the table, we report the coupling thresholds for coupling memory $m=1,2,3$, denoted by $\varepsilon_{\text{SC}}^{1}$, $\varepsilon_{\text{SC}}^{2}$, and $\varepsilon_{\text{SC}}^{3}$, respectively. The gap to the Shannon limit is shown by $\delta_{\text{SH}}=(1-R)-\varepsilon_{\text{MAP}}$.

\begin{table*}[t]
\caption{Thresholds for punctured spatially coupled turbo codes}
\begin{center}
\begin{tabular}{ccccccccccc}
\hline
Ensemble& Rate &states& $\rho_2$ & $\varepsilon_{\text{BP}}$ & $\varepsilon_{\text{MAP}}$  &$\epsSCone$ & $\epsSCthree$ & $\epsSCfive$ &$m_{\text{min}}$ &$\delta_{\text{SH}}$ \\
\hline
$\CPCCa$/$\CPCCSa$ & $1/3$ &4& 1.0 & 0.6428 & 0.6553 & 0.6553 & 0.6553 & 0.6553 &1 &0.0113\\[0.5mm]
$\CSCCa$/$\CSCCSa$ & $1/3$ &4&1.0 & 0.5405 & 0.6654 & 0.6437 & 0.6650 & 0.6654 &4 &0.0012\\[0.5mm]
$\CPCCa$/$\CPCCSa$ & $1/3$ &8& 1.0 &0.6368& 0.6621 & 0.6617 & 0.6621 & 0.6621&2&0.0045\\[0.5mm]
$\CSCCa$/$\CSCCSa$ & $1/3$ &8&1.0 &0.5026&0.6663& 0.6313&0.6647&0.6662&6&0.0003\\[0.5mm]
\hline
$\CPCCa$/$\CPCCSa$ & $1/2$ & 4&0.5 &  0.4606 & 0.4689 & 0.4689 & 0.4689 & 0.4689 &1& 0.0311\\[0.5mm]
$\CSCCa$/$\CSCCSa$ & $1/2$ & 4&0.5 & 0.3594 & 0.4981 & 0.4708 & 0.4975 & 0.4981 & 5&0.0019\\[0.5mm]
$\CPCCa$/$\CPCCSa$ & $1/2$ & 8&0.5 &0.4651 & 0.4863&0.4862&0.4863&0.4863&2&0.0137\\[0.5mm]
$\CSCCa$/$\CSCCSa$ & $1/2$ & 8&0.5 &0.3120 & 0.4993&0.4507 &0.4970&0.4992&7&0.0007\\[0.5mm]
\hline
$\CPCCa$/$\CPCCSa$ & $2/3$ & 4&0.25 &  0.2732 & 0.2772 & 0.2772 & 0.2772 & 0.2772 &1&0.0561\\[0.5mm]
$\CSCCa$/$\CSCCSa$ & $2/3$ & 4&0.25 & 0.2038 & 0.3316 & 0.3303 & 0.3305 & 0.3315 & 6&0.0018\\[0.5mm]
$\CPCCa$/$\CPCCSa$ & $2/3$ & 8&0.25 &0.2945& 0.3080&0.3080& 0.3080& 0.3080&1 &0.0253\\[0.5mm]
$\CSCCa$/$\CSCCSa$ & $2/3$ & 8&0.25 &0.1507&0.3326&0.2710&0.3278 &0.3323&7&0.0007\\[0.5mm]
\hline
$\CPCCa$/$\CPCCSa$ & $3/4$ & 4&0.166 & 0.1854 & 0.1876 & 0.1876 & 0.1876 &  0.1876 & 1&0.0624\\[0.5mm]
$\CSCCa$/$\CSCCSa$ & $3/4$ & 4&0.166 & 0.1337 & 0.2486 & 0.2155 & 0.2471 & 0.2486 & 5&0.0014\\[0.5mm]
$\CPCCa$/$\CPCCSa$ & $3/4$ & 8&0.166 &0.2103&0.2196&0.2196&0.2196&0.2196&1&0.0304\\[0.5mm]
$\CSCCa$/$\CSCCSa$ & $3/4$ & 8&0.166 &0.0865 &0.2495&0.1827&0.2416&0.2488&8&0.0005\\[0.5mm]
\hline
$\CPCCa$/$\CPCCSa$ & $4/5$ & 4&0.125 & 0.1376 & 0.1391 & 0.1391 & 0.1391 & 0.1391 & 1&0.0609\\[0.5mm]
$\CSCCa$/$\CSCCSa$ & $4/5$ & 4&0.125 & 0.0942 & 0.1990 & 0.1644 & 0.1968 & 0.1989 & 7&0.0011\\[0.5mm]
$\CPCCa$/$\CPCCSa$ & $4/5$ & 8&0.125 & 0.1628& 0.1698& 0.1698 &0.1698 &0.1698&1 &0.0302\\[0.5mm]
$\CSCCa$/$\CSCCSa$ & $4/5$ & 8&0.125 &0.0517&0.1996&0.1302 &0.1885&0.1982&8&0.0004\\[0.5mm]
\hline
$\CPCCa$/$\CPCCSa$ & $9/10$ & 4&0.055 & 0.0578 & 0.0582 & 0.0582 & 0.0582 & 0.0582 & 1&0.0418\\[0.5mm]
$\CSCCa$/$\CSCCSa$ & $9/10$ & 4&0.055 & 0.0269 & 0.0996 & 0.0624 & 0.0930 & 0.0988 & 8&0.0012\\[0.5mm]
$\CPCCa$/$\CPCCSa$ & $9/10$ & 8&0.055 &0.0732&0.0761&0.0761& 0.0761&0.0761&1&0.0239\\[0.5mm]
$\CSCCa$/$\CSCCSa$ & $9/10$ & 8&0.055 &0.0128& 0.0999 & 0.0384& 0.0765 &0.0931 &16&0.0001\\[0.5mm]
\hline

\end{tabular} 
\end{center}
\label{Tab:BPThresholdsSCC} 
\end{table*}

\begin{table*}[t]
	\caption{Thresholds for punctured Braided Convolutional Codes}
	\begin{center}
		\begin{tabular}{cccccccccc}
			\hline
			Ensemble& Rate &states& $\rho_2$ & $\varepsilon_{\text{BP}}$ & $\varepsilon_{\text{MAP}}$  &$\epsSCone$ & $\epsSCthree$ & $\epsSCfive$ & $\delta_{\text{SH}}$ \\
			\hline
			Type-I& $1/3$ &4& 1.0 & 0.5541 & 0.6653 & 0.6609&0.6644 & 0.6650& 0.0013\\[0.5mm]
			Type-II & $1/3$ &4&1.0 &0.5541 &  0.6653& 0.6651 &0.6653 &0.6653 &0.0013\\[0.5mm]
			\hline
			Type-I & $1/2$ & 4&0.5 &0.3013&0.4993&0.4932&0.4980&  0.4988&0.0007 \\[0.5mm]
			Type-II & $1/2$ & 4&0.5 &0.3013&0.4993& 0.4988&0.4993&0.4993&0.0007 \\[0.5mm]
			\hline
			Type-I & $2/3$ & 4&0.25 & -- &0.3331&0.3257&0.3315&0.3325&0.0002\\[0.5mm]
			Type-II & $2/3$ & 4&0.25 & -- &0.3331& 0.3323&0.3331 &0.3331&0.0002\\[0.5mm]
			\hline
			Type-I & $3/4$ & 4&0.166& -- &0.2491&0.2411&0.2473&0.2484&0.0009\\[0.5mm]
			Type-II & $3/4$ & 4&0.166 & -- &0.2491&0.2481&0.2491 &0.2491 &0.0009\\[0.5mm]
			\hline
			Type-I & $4/5$ & 4&0.125 & -- &0.1999&0.1915 &0.1979&0.1991 &0.0001\\[0.5mm]
			Type-II & $4/5$ & 4&0.125 & -- &0.1999&0.1986&0.1999&0.1999&0.0001\\[0.5mm]
			\hline
			Type-I & $9/10$ & 4&0.055 & -- &0.0990&0.0893&0.0966&0.0980&0.0010 \\[0.5mm]
			Type-II & $9/10$ & 4&0.055& -- &0.0990&0.0954& 0.0990&0.0990&0.0010\\[0.5mm]
			
			\hline
			
		\end{tabular} 
	\end{center}
	\label{BPThresholdsBCC} 
\end{table*}

For large enough coupling memory, we observe threshold saturation for both SC-PCCs and SC-SCCs. The value $m_{\text{min}}$ in Table~\ref{Tab:BPThresholdsSCC} denotes the smallest coupling memory for which threshold saturation is observed numerically. Interestingly, thanks to the threshold saturation phenomenon, for large enough coupling memory SC-SCCs achieve better BP threshold than SC-PCCs. We remark that SCCs yield better minimum Hamming distance than PCCs \cite{Benedetto98Serial}.

Comparing ensembles with 8-state component encoders and ensembles with 4-state component encoders, we observe that the MAP threshold improves for all the considered cases, since the overall codes become stronger. For PCCs, the BP threshold also improves for 8-state component encoders, but only with puncturing, i.e., for $R>1/3$. For SCCs, on the other hand, the BP threshold gets worse if higher memory component encoders are used. Due to this fact, a higher coupling memory $m_{\text{min}}$ is needed for SC-SCCs with 8-state component encoders until threshold saturation is observed, and this effect becomes more pronounced for larger rates. However, the achievable BP thresholds of SC-SCCs are better than those of SC-PCCs for all rates.

In Table~\ref{BPThresholdsBCC}, we give BP thresholds for Type-I and Type-II SC-BCCs with different coupling memories and several rates.\footnote{The BP threshold of the Type-I BCC with $m=1$ corresponds to the BP threshold of the original BCC.} As for PCCs, $\rho_2$ is the permeability rate of the parity bits of the upper encoder and the lower encoder. We also report the BP threshold and MAP threshold of the uncoupled ensembles.
Almost in all rates, the BP decoder works poorly for uncoupled BCCs and the BP thresholds are worse than those of PCCs and SCCs (an exception are SCCs with $R=1/3$). This is specially significant for rates $R\ge 2/3$, for which the BP thresholds of uncoupled BCCs are very close to zero.
On the other hand, the MAP thresholds of BCCs are better than those of both PCCs and SCCs for all rates. As for SC-PCCs and SC-SCCs, the BP thresholds improve if coupling is applied.
Type-II BCCs yield better thresholds than Type-I BCCs and achieve threshold saturation for small coupling memories.
In contrast, for the coupling memories considered, threshold saturation is not observed for Type-I BCCs.



For comparison purposes, in Table~\ref{BPThresholdsLDPC} we report the  $\varepsilon_{\text{BP}}$, $\varepsilon_{\text{MAP}}$, and $\varepsilon_{\text{SC}}^{1}$ for three rate-$1/2$ LDPC code ensembles.
As it is well known, by increasing the variable node degree, the MAP threshold improves, but the BP threshold decreases.
Similarly to TCs, applying the coupling improves the BP threshold.
Among all the ensembles shown in Table \ref{BPThresholdsLDPC}, the $(5,10)$ LDPC ensemble has the best MAP threshold.
However, for this ensemble the gap between the BP and MAP thresholds is larger than that of the other LDPC code ensembles and the coupling (with $m=1$) is not able to completely close this gap, therefore $\varepsilon_{\text{SC}}^{1}$ is worse than that of other two SC-LPDC code ensembles. Among all codes in Table~\ref{BPThresholdsLDPC}, the best $\varepsilon_{\text{BP}}$ is achieved by the Type II BCC ensemble. Similar to the $(5,10)$ LDPC code ensemble, the gap between the BP and the MAP threshold is relatively large for BCCs. However, for BCCs the BP threshold increases significantly after applying coupling with $m=1$.
In addition, the only way to increase the MAP threshold of the LDPC codes is to increase their variable node degree, but in TCs the BP threshold can be improved by several different methods, e.g., increasing the component code memory, selecting a good ensemble, or increasing the variable node degree.

\begin{table}[t]
	\caption{Thresholds for rate-$1/2$ TCs, SC-TCs, LDPC and SC-LDPC codes}
	\begin{center}
		\begin{tabular}{lcccc}
			\hline
			Ensemble&states&$\varepsilon_{\text{BP}}$ & $\varepsilon_{\text{MAP}}$  &$\epsSCone$ \\
			\hline
			LDPC $(3,6)$&-& 0.4294& 0.4881& 0.4880\\[0.5mm]
			LDPC $(4,8)$&-&0.3834&  0.4977& 0.4944\\[0.5mm]
			LDPC $(5,10)$&-&0.3415&0.4994& 0.4826\\[0.5mm]
			\hline
			$\CPCCa$/$\CPCCSa$&4&0.4606&0.4689&0.4689\\[0.5mm]
            $\CSCCa$/$\CSCCSa$&4&0.3594&0.4981& 0.4708\\[0.5mm]
            $\CPCCa$/$\CPCCSa$&8&0.4651&0.4863&0.4862\\[0.5mm]
            $\CSCCa$/$\CSCCSa$&8&0.3120&0.4993& 0.4507\\
            Type-I $\CBCC$&4&0.3013 &0.4993&0.4932\\[0.5mm]
            Type-II $\CBCC$&4& 0.3013&0.4993& 0.4988\\[0.5mm]
			\hline
			
		\end{tabular} 
	\end{center}
	\label{BPThresholdsLDPC} 
\end{table}

\begin{figure}[t]
	\centering
	\includegraphics[width=0.9\linewidth]{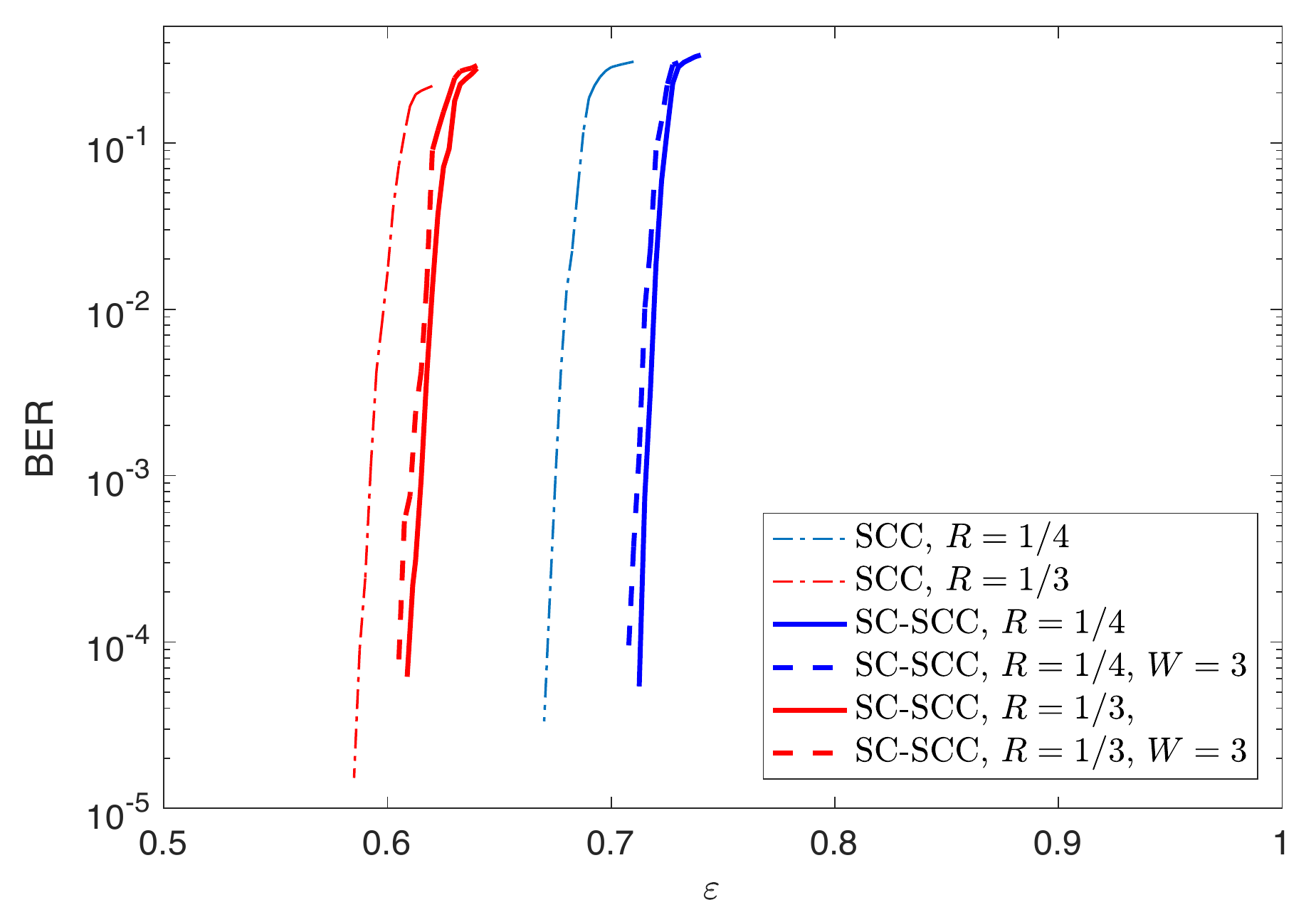}
	\caption{BER results for SC-SCCs with $L=100$ and $m=1$ on the binary erasure channel.}
	\label{SimSCC}
\end{figure}

Fig.~\ref{SimSCC} shows the bit error rate (BER) for SC-SCCs with $L=100$ and $m=1$ on the binary erasure channel for two different rates, $R=1/4$ (solid blue line) and $R=1/3$ (solid red line). 
Here, we consider the coupling of SCCs with block length $K=1024$, hence the information block length of the SC-SCC ensemble is $K=101376$.
In addition, we plot in the figure the BER curves for the uncoupled ensemble (dotted lines) with $K=3072$.
For comparison, we also plot the BER using a sliding window decoder with window size $W=3$ and $K=1024$ (dashed lines) which has a decoding latency equal to that of the uncoupled ensemble.
For both rates, the BER improves significantly applying coupling and the use of the window decoder entails only a slight performance degradation with respect to full decoder
\footnote{In this work, we are focusing on the BER of TC  and SC-TC ensembles in the waterfall region.
			 However, spatial coupling does also preserve, or even improve, the error floor performance.
			 For example, the minimum distance of each SC-TC ensemble is lower bounded by the minimum distance of the corresponding uncoupled TC ensemble. This can be shown by extending the results for BCCs derived in \cite{MoloudiISITA}.}. 
We remark that the comparison between SC-TCs and other types of codes is a new and ongoing field of research.
In \cite{zhu2015window} the authors have compared BCC and SC-LDPC codes for rate $1/2$ and under the assumption of similar latency for both.
The results in \cite{zhu2015window} show that the considered BCC ensemble outperforms the SC-LDPC code ensemble.

\section{Threshold Saturation}\label{Sec7}

The numerical results in the previous section suggest that threshold
saturation occurs for SC-TCs. In this section, for some relevant
ensembles, we prove that, indeed, threshold saturation occurs.
To prove threshold saturation we use the proof technique based on potential functions introduced in \cite{Yedla2012,Yedla2012vector}.
In the general case, the DE equations of TCs form a vector recursion.
However, we show that, for some relevant TC ensembles, it is possible to rewrite the DE vector recursion in a form which corresponds to the recursion of a scalar admissible system.
We can then prove threshold saturation using the framework in \cite{Yedla2012} for scalar recursions.
Since the proof for scalar recursions is easier to describe, we first address this case, and we then highlight the proof for the general case of TCs with a vector recursion based on the framework in \cite{Yedla2012vector}.

\begin{definition}[\cite{Yedla2012,Yedla2014}] \label{def1} A scalar admissible system $(f,g)$, is defined by the recursion
\begin{equation}
\label{recursion}
x^{(i)}=f\Big( g(x^{(i-1)});\varepsilon\Big),
\end{equation}
where $f : [0,1] \times [0,1] \rightarrow [0,1]$ and $g : [0,1] \rightarrow [0,1]$ satisfy the following conditions.
\begin{enumerate}
\item $f$ is increasing in both arguments $x,\varepsilon \in (0,1]$; 
\item $g$ is increasing in $x \in (0,1]$; 
\item $f(0;\varepsilon)=f(x;0)=g(0)=0$;
\item $f$ and $g$ have continuous second derivatives.
\end{enumerate}
\end{definition}


In the following we show that the DE equations for some relevant TCs form a scalar admissible system.

\subsection{Turbo-like codes as Scalar Admissible Systems}
\subsubsection{PCC}
The DE equations \eqref{DEPCC1}--\eqref{DEPCC6} form a vector recursion. However, if the code is built from identical component encoders, i.e., $f_{\text{U},\text{s}}=f_{\text{L},\text{s}}\triangleq f_{\text{s}}$, it follows
\[
p_{\text{U},\text{s}}^{(i)}=p_{\text{L},\text{s}}^{(i)}\triangleq x^{(i)}.
\]
Using this and substituting \eqref{DEPCC3} into \eqref{DEPCC1} and \eqref{DEPCC6} into \eqref{DEPCC4}, the DE can then be written as
\begin{equation}
\label{recursionPCC}
x^{(i)}=f_{\text{s}}(\varepsilon x^{(i-1)},\varepsilon  ),
\end{equation}
with initialization $x^{(0)}=1$.
\begin{lemma}
\label{LemmaPCC}
The DE recursion of a PCC with identical component encoders, given in \eqref{recursionPCC}, forms a scalar admissible system with $f(x;\varepsilon)=f_s(\varepsilon\cdot x,x)$ and $g(x)=x$.
\end{lemma}
\begin{IEEEproof}
 It is easy to show that all conditions in Definition~\ref{def1} are satisfied for $g(x)=x$.
We now prove that $f(x;\varepsilon)$ satisfies Conditions 1, 3 and 4. Note that $f(x;\varepsilon)$ is the transfer function of a rate-$1/2$
convolutional encoder. According to equation \eqref{eq:Transfer1},
this function can be written as $f(p_1,p_2)$, where $p_1=\varepsilon
\cdot x$ and $p_2=\varepsilon$. Using Lemma~\ref{Lemma1}, $f(p_1,p_2)$ is increasing with $p_1$ and $p_2$, therefore $f(x;\varepsilon)$ is increasing with $x$ and $\varepsilon$ and Condition 1 is satisfied.

To show that Condition 3 holds, it is enough to realize that for $\varepsilon=0$ the input sequence can be recovered perfectly
from the received sequence, i.e., $f(x;0)=0$, as there is a one-to-one mapping between input sequences and coded
sequences. Furthermore, when $x=0$, the input sequence is fully known by a-priori information and the erasure probability at the output of the decoder is zero, i.e., $f(x;0)=0$.

Finally, $f(x;\varepsilon)$ is a rational function and its poles are outside the interval $x,\varepsilon \in [0,1]$ (otherwise we may get infinite output erasure probability for a finite input erasure probability), hence it has continuous first and second derivatives inside this interval.
\end{IEEEproof} 

\subsubsection{SCC}
Consider the DE equations of the SCC ensemble in \eqref{DESCC1}--\eqref{DESCC6}, which form a vector recursion. For identical component encoders, $f_{\text{I},\text{s}}=f_{\text{O},\text{s}}\triangleq f_{\text{s}}$ and $f_{\text{I},\text{p}}=f_{\text{O},\text{p}}\triangleq f_{\text{p}}$. 
Using this and $q_{\text{I}}^{(i)}\triangleq x^{(i)}$, by substituting \eqref{DESCC2}--\eqref{DESCC6} into \eqref{DESCC1}, the DE recursion can be rewritten as 
\begin{equation}
\label{eq:SCCrec}
x^{(i)}=\varepsilon \cdot f_{\text{s}}\Big(\varepsilon g(x^{(i-1)}),\varepsilon\Big),
\end{equation}
where
\begin{equation}
\label{eq:gSCC}
g(x^{(i)})=\frac{f_{\text{s}}\Big(x^{(i)},x^{(i)}\Big)+f_{\text{p}}\Big(x^{(i)},x^{(i)}\Big)}{2},
\end{equation}
and the initial condition is $x^{(0)}=1$.

\begin{lemma}
\label{LemmaSCC}
The DE recursion of a SCC with identical component encoders, given in \eqref{eq:SCCrec} and \eqref{eq:gSCC}, form a scalar admissible system with $f(x;\varepsilon)=\varepsilon \cdot f_{\text{s}}(\varepsilon \cdot x, \varepsilon)$ and
\begin{align*}
g(x)=\frac{f_{\text{s}}(x,x)+f_{\text{p}}(x,x)}{2}.
\end{align*}
\end{lemma}
\begin{IEEEproof}
The proof follows the same arguments as the proof of Lemma~\ref{LemmaPCC}.
\end{IEEEproof}

\subsubsection{BCC}
Similarly to PCCs and SCCs, the DE equations of BCCs (see \eqref{DEBCC4}--\eqref{DEBCC6}) form a vector recursion. With identical component encoders, due to the symmetric structure of the code, $f_{\text{U},k}=f_{\text{L},k}\triangleq f_k$ and $p_{\text{U},k}^{(i)}=p_{\text{U},k}^{(i)}\triangleq x_k^{(i)}$ for $k=1,2,3$.
Using this, \eqref{DEBCC4}--\eqref{DEBCC6} can be rewritten as
\begin{align}
\label{BCC1}
x_1^{(i)}&=f_1\Big(\varepsilon \cdot x_1^{(i-1)},\varepsilon \cdot x_3^{(i-1)},\varepsilon\cdot x_2^{(i-1)}\Big)\\
\label{BCC2}
x_2^{(i)}&=f_2\Big(\varepsilon \cdot x_1^{(i-1)},\varepsilon \cdot x_3^{(i-1)},\varepsilon \cdot x_2^{(i-1)}\Big)\\
\label{BCC3}
x_3^{(i)}&=f_3\Big(\varepsilon \cdot x_1^{(i-1)},\varepsilon \cdot x_3^{(i-1)},\varepsilon \cdot x_2^{(i-1)}\Big).
\end{align}

The above DE equations are still a vector recursion.
To write the recursion in scalar form, it is necessary to have identical transfer functions for all the edges which are connected to factor nodes $T^{\text{U}}$ and $T^{\text{L}}$. This is needed because all variable nodes in a BCC receive a-priori information.
In order to achieve this property, we can apply some averaging over the different types of code symbols. In particular, we can randomly permute the order of the encoder outputs $v_\tau^{(l)}$, $l=1,\dots,n$. For each trellis section $\tau$ the order of these $n$ symbols is chosen indepently according to a uniform distribution.
Equivalently, instead of performing this permutation on the encoder outputs we can define a corresponding component encoder with a time-varying trellis in which the branch labels are permuted accordingly.
Then, it results $x_1^{(i)}=x_2^{(i)}=x_3^{(i)}\triangleq x^{(i)}$
and all transfer functions are equal to the average of the transfer functions $f_1,f_2,f_3$,
\[
f_{\text{ave}}=\frac{f_1+f_2+f_3}{3}.
\] 
Using this, the DE equations can be simplified as
\begin{equation}
\label{eq:BCCScalar}
x^{(i)}=f_{\text{ave}}(\varepsilon \cdot x^{(i-1)},\varepsilon \cdot x^{(i-1)},\varepsilon \cdot x^{(i-1)}).
\end{equation}
\begin{lemma}
\label{LemmaBCC}
The DE recursion of a BCC with identical component encoders and time varying trellises, given in \eqref{eq:BCCScalar}, form a scalar admissible system with $f(x;\varepsilon)=f_{\text{ave}}(\varepsilon \cdot x,\varepsilon \cdot x,\varepsilon \cdot x)$ and $g(x)=x$.
\end{lemma}
\begin{IEEEproof}
The proof follows the same arguments as the proof of Lemma~\ref{LemmaPCC}.
\end{IEEEproof}

\subsection{Single System Potential}
\begin{definition}[\cite{Yedla2012,Yedla2014}] \label{def2}
For a scalar admissible system, defined in Definition~\ref{def1}, the potential function $U(x;\varepsilon)$ is
\begin{align}
\label{Potential}
U(x;\varepsilon)&=\int_{0}^{x}\big{(}z-f(g(x);\varepsilon)\big{)}g'(z)dz \\
&=xg(x)-G(x)-F(g(x);\varepsilon),\nonumber
\end{align}
where $F(x;\varepsilon)=\int_{0}^{x}f(z;\varepsilon) dz$ and $G(x)=\int_{0}^{x}g(z) dz$.
\end{definition}
\begin{proposition}[\cite{Yedla2012,Yedla2014}]
The potential function has the following properties.
\begin{enumerate}
\item $U(x;\varepsilon)$ is strictly decreasing in $\varepsilon \in (0,1]$;
\item An $x\in [0,1]$ is a fixed point of the recursion
(\ref{recursion}) if and only if it is a stationary point of the corresponding potential function.
\end{enumerate}
\end{proposition}
\begin{definition}[\cite{Yedla2012,Yedla2014}] \label{defBP} If the DE recursion is the recursion of a BP decoder, the BP threshold is \cite{Yedla2012}
\[
\varepsilon^{\text{BP}}=\sup\Big\{\varepsilon
  \in[0,1] : U'(x;\varepsilon)>0,\; \forall x\in (0,1]\Big\} \ .
\] 
\end{definition}
 According to Definition~\ref{defBP},  for $\varepsilon < \varepsilon^{\text{BP}}$, the derivative of the potential function is always larger than zero for $x\in (0,1]$, i.e., the potential function has no stationary point in $x\in (0,1]$. 
\begin{definition}[\cite{Yedla2012,Yedla2014}]
\label{defPotth}
For $\varepsilon >\varepsilon^{\text{BP}} $, the minimum unstable fixed point is $u(\varepsilon)=\sup\big\{\tilde{x} \in [0,1] : f(g(x);\varepsilon)<x, x\in (0,\tilde{x})\big\}$.  Then, the potential threshold is defined as \cite{Yedla2012}
\begin{align*}
\varepsilon^*=\sup \Big\{\varepsilon \in [0,1] : u(x)>0, \min_{x \in [u(x),1]} U(x;\varepsilon)> 0 \Big\} \ .
\end{align*}
\end{definition}
The potential threshold depends on the functions $f(x;\varepsilon)$ and $g(x)$. 

\begin{figure}[t]
  \centering
    \includegraphics[width=\linewidth]{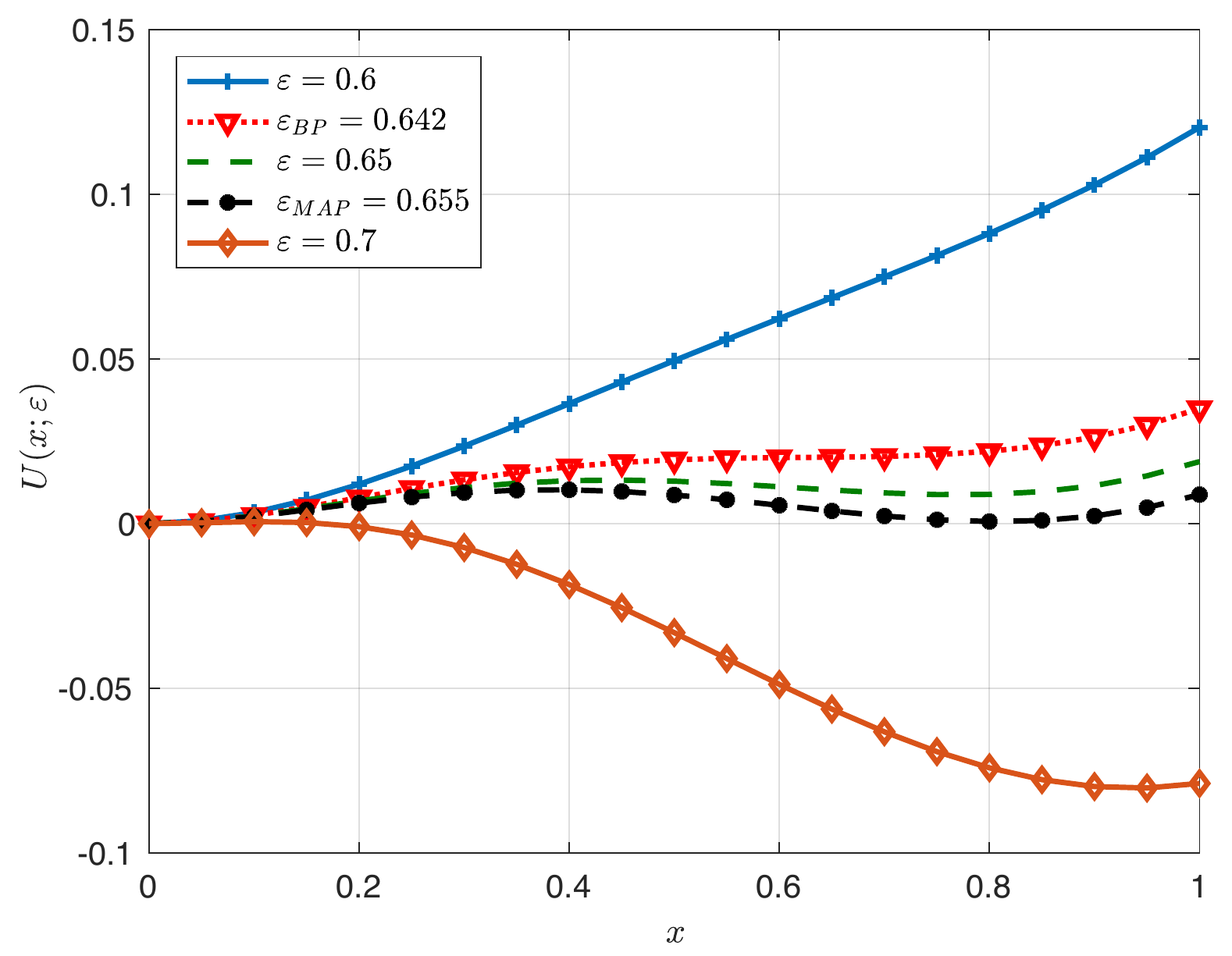}
\caption{Potential function of a PCC ensemble.}
\label{PotPCC}
\end{figure}

\begin{example}
Consider rate-$1/3$ PCCs  with identical 2-state component encoders with generator matrix ${\bs{G}}=(1,1/3)$.
For this code ensemble,
\[
f_{\text{s}}(\varepsilon\cdot x,\varepsilon)=\frac{x\varepsilon^2(2-2\varepsilon+x\varepsilon^2)}{(1-\varepsilon+x\varepsilon^2)^2} \ .
\]
Therefore,
\[
F_{\text{s}}(x;\varepsilon)=\frac{x\varepsilon^2}{1-\varepsilon+x\varepsilon^2} \ ,
\]
and
\[
U(x;\epsilon)=\frac{x\varepsilon^3+(1-\varepsilon-2\varepsilon^2)x^2}{2(1-\varepsilon+x\varepsilon^2)} \ . 
\] \hfill $\triangle$
\end{example}
\begin{example}
Consider the PCC ensemble in Fig.~\ref{Uncoupled}(b) with identical component encoders with generator matrix $\bold{G}=(1,5/7)$.
The DE recursion of this ensemble is given in \eqref{recursionPCC}, where $f_s$ is the transfer function of the $(1,5/7)$ component encoder. The corresponding potential function is
\begin{equation}
U(x;\epsilon)=x^2-G(x)-F_{\text{s}}(x;\epsilon)=\frac{x^2}{2}-F_{\text{s}}(x;\epsilon) \ ,
\end{equation}  
where $F_{\text{s}}(x;\varepsilon)=\int_{0}^{x}f_{\text{s}}(\varepsilon\cdot z,\varepsilon) dz$ and $G(x)=\int_{0}^{x}g(z)dz=\frac{x^2}{2}$. The potential function is shown in Fig.~\ref{PotPCC} for several values of $\varepsilon$.
As it is illustrated, for $\varepsilon<0.6428$ the potential function has no stationary point. The BP threshold and the potential threshold are $\varepsilon=0.6428$ and $\varepsilon=0.6553$, respectively (see Definitions~\ref{defBP} and \ref{defPotth}).
These results match with the DE results in Table~\ref{Tab:BPThresholdsSCC}. \hfill $\triangle$
\end{example}
\begin{figure}[t]
  \centering
    \includegraphics[width=\linewidth]{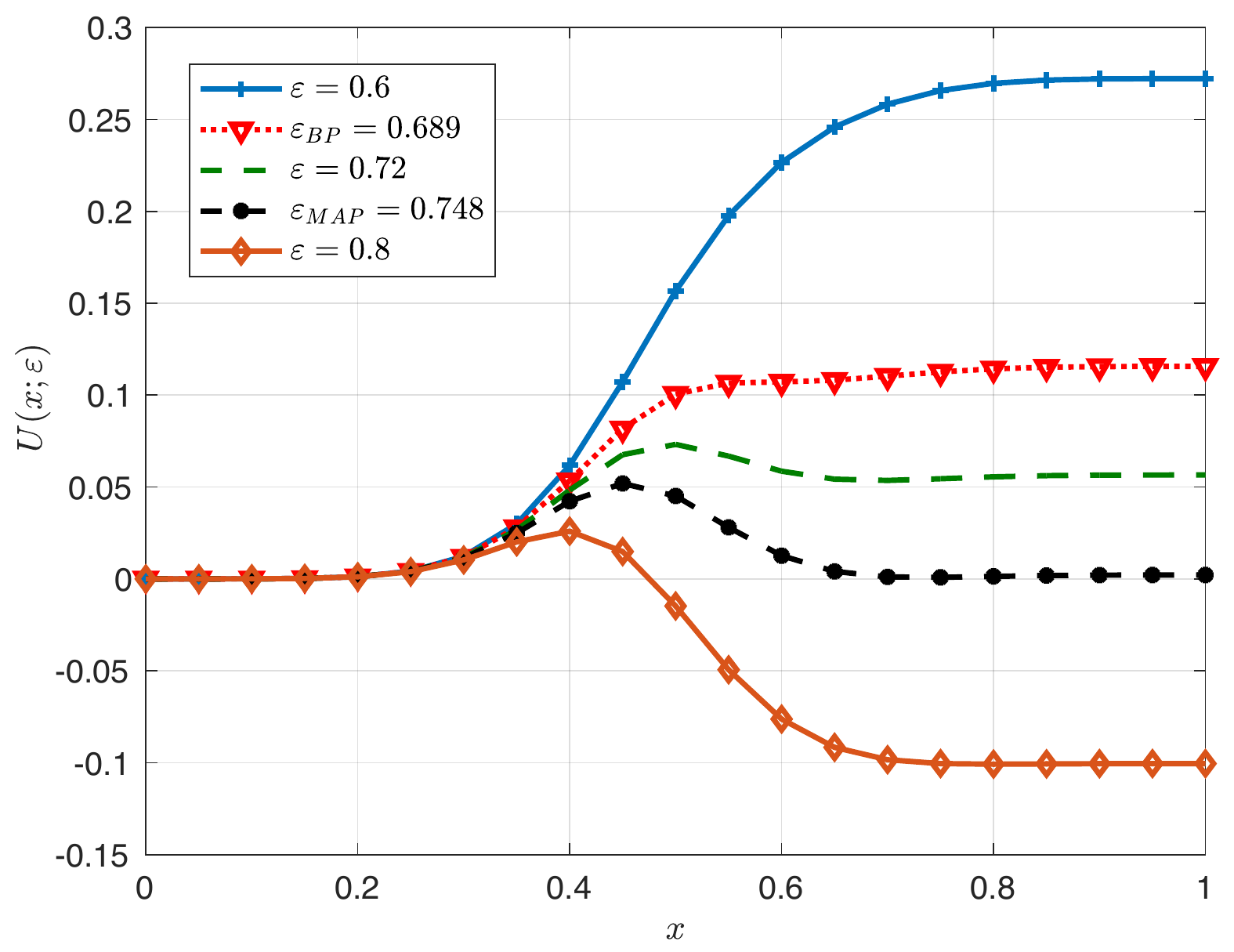}
\caption{Potential function of a SCC ensemble.}
\label{PotSCC}
\end{figure}
\begin{figure}[t]
  \centering
    \includegraphics[width=\linewidth]{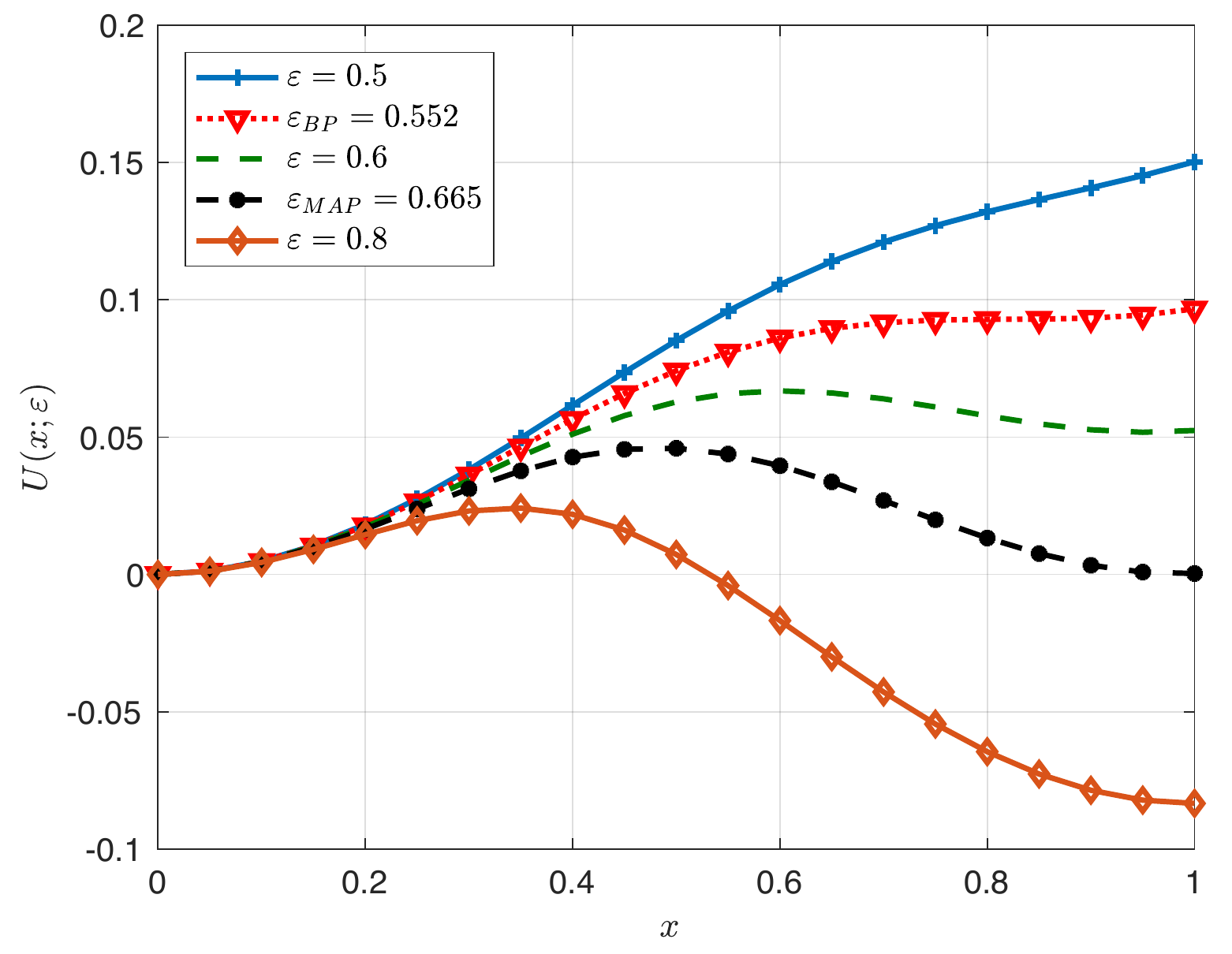}
\caption{Potential function of a BCC ensemble.}
\label{PotBCC}
\end{figure}
\begin{example} The potential function of the SCC ensemble in Fig.~\ref{Uncoupled}(c) with identical component encoders with generator matrix $\bold{G}=(1,5/7)$ is shown in Fig.~\ref{PotSCC}. The BP threshold and the potential threshold are 
$\varepsilon=0.689$ and $\varepsilon=0.748$, respectively, which match with the DE results in Table~\ref{Tab:BPThresholdsSCC}. \hfill $\triangle$
\end{example}
\begin{example} Consider the BCC ensemble in
  Fig.~\ref{Uncoupled}(d) with identical component encoders with generator matrix given in \eqref{eqG} and time-varying trellises. The potential function of this code is depicted in Fig.~\ref{PotBCC}. The BP threshold and the potential threshold are $\varepsilon=0.5522$ and $\varepsilon=0.6654$, respectively. Note that these values are slightly different from the values in
Table~\ref{BPThresholdsBCC}. This is due to the fact that we considered an ensemble with time-varying trellises, which can be modeled by means of a scalar recursion. The ensemble considered in Table~\ref{BPThresholdsBCC} needs to be analyzed by means of a vector recursion. \hfill $\triangle$
\end{example}

\subsection{Coupled System and Threshold Saturation}
\begin{theorem}\label{thm1}
Consider a spatially coupled system defined by the following recursion at time $t$,
\begin{align}
\label{SCrecursion}
x_t^{(i)}=\frac{1}{1+m}\sum_{j=0}^{m}f_{t+j}\Big(\frac{1}{1+m}\sum_{k=0}^{m}g(x_{t+j-k}^{(i-1)});\varepsilon\Big).
\end{align}
If $f(x;\varepsilon)$ and $g(x)$ form a scalar admissible system, for large enough coupling memory and $\varepsilon < \varepsilon^*$, the only fixed point of the recursion is
$x=0$.
\end{theorem}
\begin{IEEEproof}
The proof follows from \cite{Yedla2012}.
\end{IEEEproof}

In the following we show that the DE recursions of SC-TCs (with identical component encoders) can be written in the form \eqref{SCrecursion}. As a result, threshold saturation occurs for these ensembles.

\subsubsection{PCCs}
Consider the SC-PCC ensemble in Fig.~\ref{Coupled}(a) with identical component encoders.
Due to the symmetric coupling structure, it follows that (cf. \eqref{DESCPCC1} and \eqref{DESCPCC2})
\[
\bar{q}_{\text{U}}^{(i,t)}=\bar{q}_{\text{L}}^{(i,t)}\triangleq x_{t}^{(i)}.
\]
Now, using $x_t^{(i)}$ in \eqref{DESCPCC3} and \eqref{DESCPCC6}, we can write
\begin{align}
\label{eq:QLi}
q_{\text{L}}^{(i,t)}=q_{\text{U}}^{(i,t)}=\varepsilon \cdot \frac{1}{m+1}\sum_{k=0}^{m}x_{t-k}^{(i-1)}.
\end{align}
Finally, by substituting \eqref{eq:QLi} into \eqref{DESCPCC4} and \eqref{DESCPCC5} and the results into \eqref{DESCPCC1} and \eqref{DESCPCC2}, the recursion of SC-PCCs can be rewritten as
\begin{equation}
\label{SCPCCrecursion}
x_t^{(i)}=\frac{1}{1+m}\sum_{j=0}^{m}f_{\text{s},t+j}\Big(\frac{\varepsilon}{m+1}\cdot\sum_{k=0}^{m}x_{t+j-k}^{(i-1)},\varepsilon\Big).
\end{equation}
Note that the recursion in \eqref{SCPCCrecursion} is identical to the recursion in \eqref{SCrecursion}.

\subsubsection{SCCs}

Consider the SC-SCC ensemble in Fig.~\ref{Coupled}(b) with identical component encoders. Define $x_t^{(i)} \triangleq q_{\text{I}}^{(i,t)}$ (see \eqref{DESCSCC2})
Now, use it in \eqref{DESCSCC3}--\eqref{DESCSCC6}.
Finally, by substituting the result in \eqref{DESCSCC2}, the recursion of a SC-SCC an be rewritten as
\begin{equation}
\label{SCSCCrecursion}
x_t^{(i)}=\frac{1}{1+m}\sum_{j=0}^{m}\varepsilon \cdot f_{\text{s},t+j}\Big(\frac{\varepsilon}{m+1}\cdot\sum_{k=0}^{m}g(x_{t+j-k}^{(i-1)}),\varepsilon\Big),
\end{equation}
where $g(x)$ is shown in equation \eqref{eq:gSCC}. The recursion in \eqref{SCSCCrecursion} is identical to the recursion in Theorem \ref{thm1}. 

\subsubsection{BCCs}

Consider a coupling for BCCs slightly different from the one for Type-II BCCs.
At time $t$, each of the parity sequences $\bs{v}^{\text{U}}_t$ and $\bs{v}^{\text{L}}_t$ is divided into $m+1$ sequences, $\boldsymbol{v}_{t,j}^{\text{U}}$, $j=0,\dots,m$, and $\boldsymbol{v}_{t,j}^{\text{L}}$, $j=0,\dots,m$, respectively (in Type-II BCCs they are divided into $m$ sequences).
The sequences $\boldsymbol{v}_{t-j,j}^{\text{U}}$ and $\boldsymbol{v}_{t-j,j}^{\text{L}}$ are multiplexed and reordered, and are used as the second input of the lower and upper encoder, respectively. Note that in this way of coupling, part of the parity bits at time $t$ are used as input at the same time instant $t$. Now, similarly to uncoupled BCCs, consider identical time-varying trellises. Let $x^{(i)}_t$ denote the extrinsic erasure probability from $T^{\text{U}}_t$ through all its edges in the $i$th iteration. The erasure probabilities to $\text{T}^{\text{U}}_t$ through all its incoming edges are equal and are given by the average of the erasure probabilities from variable nodes $\bs{v}_{t'}$, $t'= t-m,\dots,t$, 
\[
q_t^{(i)}=\frac{\varepsilon}{1+m}\sum_{k=0}^{m}x_{t-k}^{(i-1)}.
\]
Thus, the erasure probabilities from $T^{\text{U}}_t$ and $T^{\text{L}}_t$ are identical and equal to $f_{\text{ave},t}(q_t^{(i)},q_t^{(i)},q_t^{(i)})$. Finally, the recursion at time slot $t$ is
\begin{equation}
\label{eq:SCBCC}
x_t^{(i)}=\frac{1}{1+m}\sum_{j=0}^{m}f_{\text{ave},t+j}(q_{t+j}^{(i)},q_{t+j}^{(i)},q_{t+j}^{(i)}).
\end{equation}
The recursion in (\ref{eq:SCBCC}) is identical to (\ref{SCrecursion}).

\subsection{Random Puncturing and Scalar Admissible System}

In the following, we show that the DE recursion of punctured TC ensembles can also be rewritten as a scalar admissible system for some particular cases. Then, threshold saturation follows from the discussion in the previous subsection.
\subsubsection{PCC} Consider the PCC ensemble with
identical component encoders and random puncturing of the parity bits with
permeability rate $\rho$. The DE recursion can be rewritten
as,
\[
x^{(i)}=f_{\text{s}}(\varepsilon x^{(i-1)},1-(1-\varepsilon)\rho).
\]
The above equation is a recursion of a scalar admissible system and satisfies the
conditions in Definition \ref{def1}, where $g(x)=x$ and
$f(x;\varepsilon)=f_s(\varepsilon\cdot x, 1-(1-\varepsilon)\rho)$.

\subsubsection{SCC} 

Consider random puncturing of the SCC ensemble
with identical component encoders. Assuming $\rho_0=\rho_1$ (i.e., we puncture also systematic bits of the outer code),
we can rewrite the DE recursion as
\[
x^{(i)}=\varepsilon_{\rho_1} \cdot f_{\text{s}}(\varepsilon_{\rho_1} x^{(i-1)},\varepsilon_{\rho_2}),
\]
where $\varepsilon_{\rho_1}=1-(1-\varepsilon)\rho_1$ and
$\varepsilon_{\rho_2}=1-(1-\varepsilon)\rho_2$. The above equation is
the recursion of a scalar admissible system, where
$f(x;\varepsilon)=\varepsilon_{\rho_1} f_s(\varepsilon_{\rho_1}\cdot
x,\varepsilon_{\rho_2} )$ and $g(x)$ is obtained by equation \eqref{eq:gSCC}.

\subsubsection{BCC} 

Consider random puncturing of the BCC ensemble with identical
time-varying trellises. Assume that the systematic bits and the parity bits of the upper and lower encoders are punctured with the same permeability rate $\rho$. Then, the DE recursion can be
rewritten as \eqref{eq:BCCScalar}, where $\varepsilon$ should
be replaced by $\varepsilon_{\rho}=1-(1-\varepsilon)\rho$.

\subsection{Turbo-like Codes as Vector Admissible Systems}
In general, the DE recursions of TCs are vector recursions. In this case, it is
possible to prove threshold saturation using the technique proposed in \cite{Yedla2012vector} for vector recursions. The proof is similar to that of scalar recursions, albeit more involved. In the following, we show how to rewrite the recursion of punctured PCCs as a vector admissible system recursion. Then, following \cite{Yedla2012vector}, we can prove threshold saturation. 
Using the same technique, it is possible to prove threshold saturation for SCCs and BCCs as well.


Consider the DE equations of the PCC ensemble in \eqref{DEPCC1}--\eqref{DEPCC6}. To reduce the number of the equations, substitute 
\eqref{DEPCC3} and \eqref{DEPCC6} into \eqref{DEPCC1} and \eqref{DEPCC4},
respectively.
Consider random puncturing of information bits, upper encoder parity bits and lower encoder parity bits with permeability rates
$\rho_0$, $\rho_1$ and $\rho_2$, respectively. By considering
$x_1^{(i)}\triangleq p_{\text{U,s}} $ and $x_2^{(i)}\triangleq
p_{\text{L,s}} $,  the DE recursion can be simplified to 
\[
x_1^{(i)}=f_{\text{U,s}}(\varepsilon_{\rho_0}\cdot x_2^{(i-1)},\varepsilon_{\rho_1})
\]
\[
x_2^{(i)}=f_{\text{L,s}}(\varepsilon_{\rho_0}\cdot x_1^{(i-1)},\varepsilon_{\rho_2}).
\]
The above equations can be written in vector format as
\begin{equation}
\label{PCCvector}
\bs{x}^{(i)}=\bs{f}(\bs{g}(\bs{x}^{(i-1)});\varepsilon),
\end{equation}
where, $\bs{x}=[x_1, x_2]$, $\bs{f}(\bs{x};\varepsilon)=[f_{U,\text{s}}(\varepsilon_{\rho_0} \cdot x_1,\varepsilon_{\rho_1}),f _{L,\text{s}}(\varepsilon_{\rho_0} \cdot x_2,
\varepsilon_{\rho_2})]$ and $\bs{g}(\bs{x})=[x_2,x_1]$.
Is it easy to verify that the recursion in \eqref{PCCvector} satisfies the
conditions in \cite[Def.~1]{Yedla2012vector}, hence \eqref{PCCvector} is the recursion of a vector
admissible system. 
For this vector admissible system,
the line integral is path independent in \cite[Eq.~(2)]{Yedla2012vector} and the potential function is well defined.
So, we can define (see \cite{Yedla2012vector}) $\bs{D}=I_{2\times 2}$, $G=x_1\cdot x_2$ and
\[
F=\int_0^{x_1} f_{\text{U,s}}(\varepsilon_{\rho_0} \cdot
z,\varepsilon_{\rho_1}) \; dz+\int_0^{x_2} f_{\text{L,s}}(\varepsilon_{\rho_0} \cdot z,\varepsilon_{\rho_2}) \; dz.
\]
It is possible to show that the DE recursion of SC-PCCs can be
rewritten in the same form as \cite[Eq.~(5)]{Yedla2012vector} and by
using \cite[Th.~1]{Yedla2012vector}, threshold saturation can be proven.
\section{Conclusion}\label{Sec8}
In this paper we investigated the impact of spatial coupling on the BP decoding threshold of turbo-like codes. We introduced the concept of spatial coupling for PCCs  and SCCs, and generalized the concept of coupling for BCCs.
Considering transmission over the BEC, we derived the exact DE equations for uncoupled and coupled ensembles.  
For all spatially coupled ensembles, the BP threshold improves and our numerical results suggest that threshold saturation occurs if the coupling memory is chosen sufficiently large. We therefore constructed rate-compatible families of SC-TCs that achieve close-to-capacity performance for a wide range of code rates.

We showed that the DE equations of SC-TC ensembles with identical component encoders can be properly rewritten as a scalar recursion. 
For SC-PCCs, SC-SCCs and BCCs we then proved threshold saturation analytically, using the proof technique based on potential functions proposed in \cite{Yedla2012,Yedla2014}. Finally, we demonstrated how vector recursions can be used to extend the proof to more general ensembles.

A generalization of our results to general binary-input memoryless channels is challenging, because the transfer functions of the component decoders can no longer be obtained in closed form. Even a numerical computation of the exact thresholds is difficult, but Monte Carlo methods and Gaussian approximation techniques could be helpful tools for finding approximate thresholds. EXIT charts, for example, have been widely used for analyzing uncoupled TCs  and may be useful for estimating the thresholds of SC-TCs. A connection between EXIT functions and potential functions of spatially coupled systems is given in \cite{KudekarWaveLike}. An investigation of SC-TC ensembles along this line may be an interesting direction for future work. The simulation results for SC-BCCs over the AWGN channel in \cite{ZhangBCC} and \cite{zhu2015window} clearly show that spatial coupling significantly improves the performance, suggesting that threshold saturation also occurs for this channel.

The invention of turbo codes and the rediscovery of LDPC codes, allowed to approach capacity with practical codes.
Today, both turbo-like codes and LDPC codes are ubiquitous in communication standards.
In the academic arena, however, the interest on turbo-like codes has been declining in the last years in favor of the (considered) more mathematically-appealing LDPC codes.
The invention of spatially coupled LDPC codes has exacerbated this situation.
Without spatial coupling, it is well known that PCCs yield good BP thresholds but poor error floors, while SCCs and BCCs show low error floors but poor BP thresholds.
Our SC-TCs, however, demonstrate that turbo-like codes can also greatly benefit from spatial coupling.
The concept of spatial coupling opens some new degrees of freedom in the design of codes on graphs: designing a concatenated coding scheme for achieving the best BP threshold in the uncoupled case may not necessarily lead to the best overall performance.
Instead of optimizing the component encoder characteristics for BP decoding, it is possible to optimize the MAP decoding threshold and rely on the threshold saturation effect of spatial coupling.
Powerful code ensembles with strong distance properties such as SCCs and BCCs can then perform close to capacity with low-complexity iterative decoding.
We hope that our work on spatially coupled turbo-like codes will trigger some new interest in turbo-like coding structures.



\balance
\begin{IEEEbiographynophoto}{Saeedeh Moloudi}
received the Master Degree in Wireless Communications from Shiraz University, Iran in 2012. Since September 2013, she has been a PhD candidate at the Department of Electrical and Information Technology, Lund University. Her main research interests include design and analysis of coding systems and graph based iterative algorithms.
\end{IEEEbiographynophoto}

\begin{IEEEbiographynophoto}{Michael Lentmaier}
 received the Dipl. Ing. degree in electrical engineering from University of Ulm, Germany in 1998, and the Ph.D. degree in telecommunication theory from Lund University, Sweden in 2003. He then worked as a Post-Doctoral Research Associate at University of Notre Dame, Indiana and at University of Ulm.  From 2005 to 2007 he was with the Institute of Communications and Navigation of the German Aerospace Center (DLR) in Oberpfaffenhofen, where he worked on signal processing techniques in satellite navigation receivers. From 2008 to 2012 he was a senior researcher and lecturer at the Vodafone Chair Mobile Communications Systems at TU Dresden, where he was heading the Algorithms and Coding research group.  Since January 2013 he is an Associate Professor at the Department of Electrical and Information Technology at Lund University. His research interests include design and analysis of coding systems, graph based iterative algorithms and Bayesian methods applied to decoding, detection and estimation in communication systems.  He is a senior member of the IEEE and served as an editor for IEEE Communications Letters (2010-2013), IEEE Transactions on Communications (2014-2017), and IEEE Transactions on Information Theory (since April 2017).  He was awarded the Communications Society and Information Theory Society Joint Paper Award (2012) for his paper “Iterative Decoding Threshold Analysis for LDPC Convolutional Codes”.
\end{IEEEbiographynophoto}

\begin{IEEEbiographynophoto}{Alexandre Graell i Amat}
received the MSc degree in Telecommunications Engineering from the Universitat Politècnica de Catalunya, Barcelona, Catalonia, Spain, in 2001, and the MSc and the PhD degrees in Electrical Engineering from the Politecnico di Torino, Turin, Italy, in 2000 and 2004, respectively. From September 2001 to May 2002, he was a Visiting Scholar at the University of California San Diego, La Jolla, CA. From September 2002 to May 2003, he held a Visiting Appointment at the Universitat Pompeu Fabra and at the Telecommunications Technological Center of Catalonia, both in Barcelona. From 2001 to 2004, he also held a part-time appointment at STMicroelectronics Data Storage Division, Milan, Italy, as a Consultant on coding for magnetic recording channels. From 2004 to 2005, he was a Visiting Professor at the Universitat Pompeu Fabra, Barcelona, Spain. From 2006 to 2010, he was with the Department of Electronics, Telecom Bretagne (former ENST Bretagne), Brest, France. In 2011, he joined the Department of Electrical Engineering, Chalmers University of Technology, Gothenburg, Sweden, where he is currently a Professor. His research interests include the areas of modern coding theory, distributed storage, and optical communications.
Prof. Graell i Amat is currently Editor-at-Large of the IEEE TRANSACTIONS ON COMMUNICATIONS. He was an Associate Editor of the IEEE TRANSACTIONS ON COMMUNICATIONS from 2011 to 2015, and the IEEE COMMUNICATIONS LETTERS from 2011 to 2013. He was the General Co-Chair of the 7th International Symposium on Turbo Codes and Iterative Information Processing, Gothenburg, Sweden, 2012. He received the postdoctoral Juan de la Cierva Fellowship from the Spanish Ministry of Education and Science and the Marie Curie Intra-European Fellowship from the European Commission. He received the IEEE Communications Society 2010 Europe, Middle East, and Africa Region Outstanding Young Researcher Award.
\end{IEEEbiographynophoto}
\end{document}